\documentstyle{mn}
\newread\epsffilein    % file to \read
\newif\ifepsffileok    % continue looking for the bounding box?
\newif\ifepsfbbfound   % success?
\newif\ifepsfverbose   % report what you're making?
\newdimen\epsfxsize    % horizontal size after scaling
\newdimen\epsfysize    % vertical size after scaling
\newdimen\epsftsize    % horizontal size before scaling
\newdimen\epsfrsize    % vertical size before scaling
\newdimen\epsftmp      % register for arithmetic manipulation
\newdimen\pspoints     % conversion factor
\def\epsfbox#1{\global\def\epsfllx{72}\global\def\epsflly{72}%
   \global\def\epsfurx{540}\global\def\epsfury{720}%
   \def\lbracket{[}\def\testit{#1}\ifx\testit\lbracket
   \let\next=\epsfgetlitbb\else\let\next=\epsfnormal\fi\next{#1}}%
\def\epsfgetlitbb#1#2 #3 #4 #5]#6{\epsfgrab #2 #3 #4 #5 .\\%
   \epsfsetgraph{#6}}%
\def\epsfnormal#1{\epsfgetbb{#1}\epsfsetgraph{#1}}%
\def\epsfgetbb#1{%
\openin\epsffilein=#1
\ifeof\epsffilein\errmessage{I couldn't open #1, will ignore it}\else
   {\epsffileoktrue \chardef\other=12
    \def\do##1{\catcode`##1=\other}\dospecials \catcode`\ =10
    \loop
       \read\epsffilein to \epsffileline
       \ifeof\epsffilein\epsffileokfalse\else
          \expandafter\epsfaux\epsffileline:. \\%
       \fi
   \ifepsffileok\repeat
   \ifepsfbbfound\else
    \ifepsfverbose\message{No bounding box comment in #1; using defaults}\fi\fi
   }\closein\epsffilein\fi}%
\def\epsfsetgraph#1{%
   \epsfrsize=\epsfury\pspoints
   \advance\epsfrsize by-\epsflly\pspoints
   \epsftsize=\epsfurx\pspoints
   \advance\epsftsize by-\epsfllx\pspoints
   \epsfxsize\epsfsize\epsftsize\epsfrsize
   \ifnum\epsfxsize=0 \ifnum\epsfysize=0
      \epsfxsize=\epsftsize \epsfysize=\epsfrsize
     \else\epsftmp=\epsftsize \divide\epsftmp\epsfrsize
       \epsfxsize=\epsfysize \multiply\epsfxsize\epsftmp
       \multiply\epsftmp\epsfrsize \advance\epsftsize-\epsftmp
       \epsftmp=\epsfysize
       \loop \advance\epsftsize\epsftsize \divide\epsftmp 2
       \ifnum\epsftmp>0
          \ifnum\epsftsize<\epsfrsize\else
             \advance\epsftsize-\epsfrsize \advance\epsfxsize\epsftmp \fi
       \repeat
     \fi
   \else\epsftmp=\epsfrsize \divide\epsftmp\epsftsize
     \epsfysize=\epsfxsize \multiply\epsfysize\epsftmp   
     \multiply\epsftmp\epsftsize \advance\epsfrsize-\epsftmp
     \epsftmp=\epsfxsize
     \loop \advance\epsfrsize\epsfrsize \divide\epsftmp 2
     \ifnum\epsftmp>0
        \ifnum\epsfrsize<\epsftsize\else
           \advance\epsfrsize-\epsftsize \advance\epsfysize\epsftmp \fi
     \repeat     
   \fi
   \ifepsfverbose\message{#1: width=\the\epsfxsize, height=\the\epsfysize}\fi
   \epsftmp=10\epsfxsize \divide\epsftmp\pspoints
   \newcount\figskipcount
      \message{#1 \the\epsfysize  }
   \vbox to\epsfysize{\vfil\hbox to\epsfxsize{%
      \includegraphics{#1}%
      \hfil}}%
\epsfxsize=0pt\epsfysize=0pt}%

{\catcode`\%=12 \global\let\epsfpercent=%\global\def\epsfbblit{%BoundingBox}}%
\long\def\epsfaux#1#2:#3\\{\ifx#1\epsfpercent
   \def\testit{#2}\ifx\testit\epsfbblit
      \epsfgrab #3 . . . \\%
      \epsffileokfalse
      \global\epsfbbfoundtrue
   \fi\else\ifx#1\par\else\epsffileokfalse\fi\fi}%
\def\epsfgrab #1 #2 #3 #4 #5\\{%
   \global\def\epsfllx{#1}\ifx\epsfllx\empty
      \epsfgrab #2 #3 #4 #5 .\\\else
   \global\def\epsflly{#2}%
   \global\def\epsfurx{#3}\global\def\epsfury{#4}\fi}%
\def\epsfsize#1#2{\epsfxsize}
\def\figinsert#1#2{\epsfbox{#1} \message{#2} }    %insert figures
\def \veldisp {< w^{2} >^{1/2} }

\title[The Durham/UKST Galaxy Redshift Survey]
      {The Durham/UKST Galaxy Redshift Survey - IV.\\ Redshift Space
Distortions in the 2-Point Correlation Function.}
\author[A. Ratcliffe et al.]
       {A. Ratcliffe$^{1}$, T. Shanks$^{1}$, R. Fong$^{1}$ and Q.A.
Parker$^{2}$ \\
        $^{1}$Physics Deptartment, University of Durham, South Road, Durham,
DH1 3LE.\\
	$^{2}$Anglo-Australian Observatory, Coonabarabran, NSW 2357,
Australia.}
\begin{document}

\maketitle

\begin{abstract}
We have investigated the redshift space distortions in the optically
selected Durham/UKST Galaxy Redshift Survey using the 2-point galaxy
correlation function perpendicular and parallel to the observer's line
of sight, $\xi(\sigma,\pi)$. We present results for the real space
2-point correlation function, $\xi(r)$, by inverting the optimally
estimated projected correlation function, which is obtained by
integration of $\xi(\sigma,\pi)$, and find good agreement with other
real space estimates. On small, non-linear scales we observe an
elongation of the constant $\xi(\sigma,\pi)$ contours in the line of
sight direction. This is due to the galaxy velocity dispersion and is
the common ``Finger of God'' effect seen in redshift surveys. Our
result for the one-dimensional pairwise rms velocity dispersion is
$\veldisp = 416 \pm 36$kms$^{-1}$ which is consistent with those from
recent redshift surveys and canonical values, but inconsistent with
SCDM or LCDM models. On larger, linear scales we observe a compression
of the $\xi(\sigma,\pi)$ contours in the line of sight direction. This
is due to the infall of galaxies into overdense regions and the
Durham/UKST data favours a value of $(\Omega^{0.6}/b)$$\sim$0.5, where
$\Omega$ is the mean mass density of the Universe and $b$ is the linear
bias factor which relates the galaxy and mass distributions. Comparison
with other optical estimates yield consistent results, with the
conclusion that the data does not favour an unbiased critical-density
universe.
\end{abstract}


\section{Introduction}

The use of redshifts as distance estimates is commonplace in surveys
which map the galaxy distribution of the Universe. While redshifts are
both quick and easy to acquire, they do not generally reflect
the true distance to the galaxy because of the galaxy's own peculiar
velocity with respect to the Hubble flow. Therefore, the observed
clustering pattern will be imprinted with the galaxy peculiar velocity
field along the observer's line of sight direction. While at first this
would appear to be highly problematic for the use and understanding of
the measured clustering statistics, these so-called redshift space
distortions can be used to estimate some important cosmological
parameters describing the dynamics of the Universe (e.g. Peebles 1980;
Kaiser 1987).

We will be investigating these redshift space distortions using the
spatial 2-point correlation function, $\xi$, and denote a real space
separation by $r$ and a redshift space separation by $s$. Non-linear
and linear effects will be competing on all scales but for the most
part we take the simplifying assumption that non-linear effects
dominate on small scales ($< 10h^{-1}$Mpc), whereas linear effects
dominate on larger scales ($> 10h^{-1}$Mpc). Although this is a
slightly naive approach we use N-body simulations and
mock catalogues to guide us in determining the accuracy of the
modelling involved.

The initial clustering results, redshift maps, etc. of the Durham/UKST
Galaxy Redshift Survey were summarized in the first paper of this
series \cite{mymnras}. In this paper we present a detailed analysis of
the redshift space distortions in the 2-point correlation function as
estimated from this optically selected survey. We briefly describe our
survey in Section~\ref{redsursec}. Section~\ref{xispsec} gives a
qualitative description of the effects of the redshift space
distortions on the 2-point correlation function. In
Section~\ref{xirsec} we describe and test methods of obtaining the real
space correlation function from redshift space estimates. In
Section~\ref{nonlinsec} we describe non-linear effects and estimate the
one-dimensional rms galaxy pairwise velocity disperison. Linear infall
effects are described in Section~\ref{linsec} and an estimate of
$\Omega^{0.6}/b$ is obtained. The results of this analysis are
discussed and compared with other redshift surveys and models of
structure formation in Section~\ref{discusssec}. Finally, we summarize
our conclusions in Section~\ref{concsec}.

\section{The Durham/UKST Galaxy Redshift Survey} \label{redsursec}

The Durham/UKST Galaxy Redshift Survey was constructed using the FLAIR
fibre optic system \cite{FLAIR} on the 1.2m UK Schmidt Telescope at
Siding Spring, Australia. This survey uses the astrometry and
photometry from the Edinburgh/Durham Southern Galaxy Catalogue (EDSGC;
Collins, Heydon-Dumbleton \& MacGillivray 1988; Collins, Nichol \&
Lumsden 1992) and was completed in 1995 after a 3-yr observing
programme. The survey itself covers a $\sim$$20^{\circ} \times
75^{\circ}$ area centered on the South Galactic Pole (60 UKST plates)
and is sparse sampled at a rate of one in three of the galaxies to
$b_{J} \simeq 17$ mag. The resulting survey contains $\sim$2500
redshifts, probes to a depth greater than $300h^{-1}$Mpc, with a median
depth of $\sim$$150h^{-1}$Mpc, and surveys a volume of space $\sim$$4
\times 10^{6} (h^{-1}$Mpc$)^{3}$.

The survey is $>$75 per cent complete to the nominal magnitude limit of
$b_{J} = 17.0$ mag. This incompleteness was mainly caused by poor
observing conditions, intrinsically low throughput fibres and other
various observational effects. In a comparison with $\sim$150 published
galaxy velocities (Peterson et al. 1986; Fairall \& Jones 1988;
Metcalfe et al. 1989; da Costa et al. 1991) our measured redshifts had
negligible offset and were accurate to $\pm 150$ kms$^{-1}$. The
scatter in the EDSGC magnitudes has been estimated at $\pm 0.22$ mags
\cite{nmb} for a sample of $\sim$100 galaxies. This scatter has been
confirmed by a preliminary analysis of a larger sample of high quality
CCD photometry. All of these observational details are discussed
further in a forthcoming data paper (Ratcliffe et al., in
preparation).

\section{Redshift Space Distortions} \label{xispsec}

We investigate the effects of redshift space distortions by estimating
the spatial 2-point correlation function, $\xi$, as a function of the
two variables $\sigma$ and $\pi$. These are the separations
perpendicular ($\sigma$) and parallel ($\pi$) to the line of sight. The
specific definitions of $\sigma$ and $\pi$ we use were given in
Ratcliffe et al. (1996c), hereafter Paper III. However, we found that
our results do not depend significantly on their exact form.

Galaxy peculiar velocities cause an otherwise isotropic real space
correlation function to become anisotropic when observed in redshift
space. The degree of anisotropy measures the low-order moments of the
peculiar velocity distribution function (e.g. Peebles 1980). On small,
non-linear scales ($< 10h^{-1}$Mpc) the velocity dispersion of galaxies
in virialised regions dominates the anisotropy, causing an elongation
in the contours of constant $\xi$ along the line of sight direction.
This is the well-known ``Finger of God'' effect seen in redshift
surveys and allows one to estimate the one-dimensional pairwise rms
velocity dispersion of galaxies, $\veldisp$ (e.g. Davis \& Peebles
1983). On larger, more linear scales ($> 10h^{-1}$Mpc) the infall of
galaxies into overdense regions dominates, causing a compression of the
$\xi$ contours in the line of sight direction (Kaiser 1987). This
effect allows a measurement of $\Omega^{0.6}/b$. While our division of
the non-linear and linear regimes is slightly naive, we will be guided
by CDM mock catalogues to determine the accuracy of the modelling in
the different regions.

\subsection{The Redshift Space 2-Point Correlation Function,
$\xi(\sigma,\pi)$} \label{xisp2sec}

We have estimated $\xi(\sigma,\pi)$ using the optimal methods
determined empirically in Paper III. Briefly, this involves
distributing a random and homogeneous catalogue with the same angular
and radial selection functions as the original survey. One then cross
correlates data-data, data-random and random-random pairs, binning them
as a function of separation (in this case $\sigma$ and $\pi$). As a
result of testing Monte Carlo mock catalogues drawn from CDM $N$-body
simulations we found that the estimator and weighting combination which
most accurately traced the actual $\xi$ and also produced the minimum
variance in $\xi$ was the estimator of Hamilton (1993) and the
weighting of Efstathiou (1988). All of these techniques and the biases
in them are discussed at length in Paper III and we refer the
interested reader there.

In Figs.~\ref{xispcdmfig} and~\ref{xispdurfig} we show contour plots of
constant $\xi$ as a function of $\sigma$ and $\pi$. On both of these
figures we adopt the following conventions: solid lines denote $\xi >
1$ with $\Delta\xi = 1$; short-dashed lines denote $0 < \xi < 1$ with
$\Delta\xi = 0.1$; and long-dashed lines denote $\xi < 0$ with
$\Delta\xi = 0.1$. For reference, the contours $\xi = 1$ and 0 are in
thick bold and the regularly spaced thin bold lines show an isotropic
correlation function for comparison. When the plotted scale is log-log
the binning size is $0.2 dex$, when it is linear-linear the binning is
$1.0h^{-1}$Mpc. In both cases no formal smoothing has been applied.

\subsection{Results from the CDM models}

\begin{figure}
\centering
\centerline{\epsfxsize=8.5truecm \figinsert{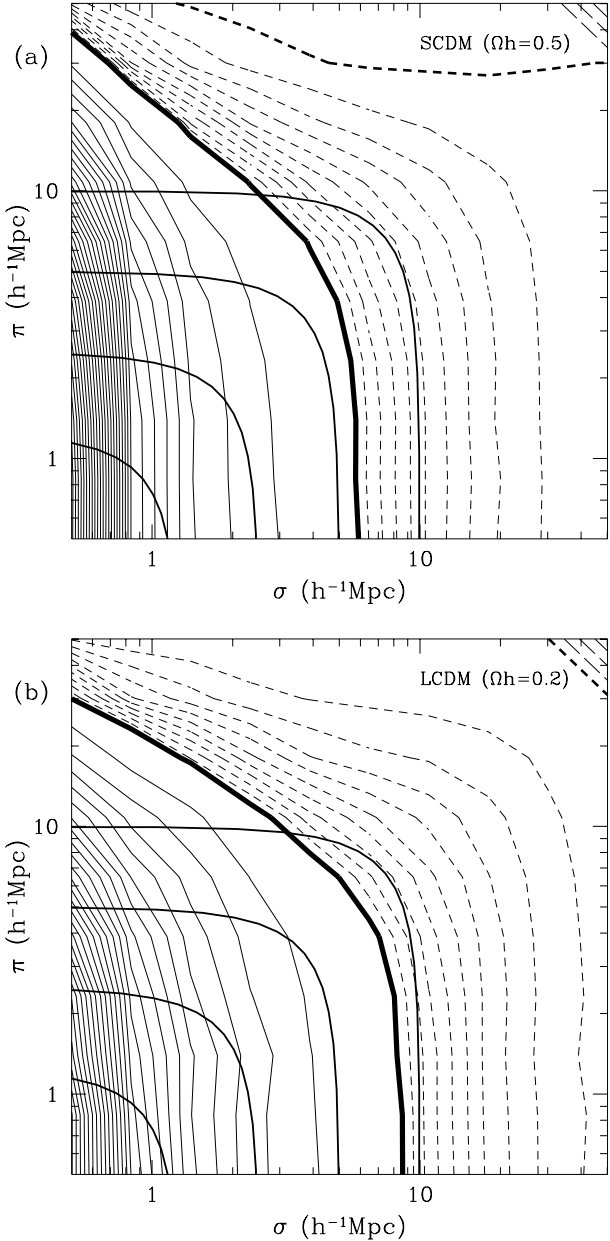}{0.0pt}}
\caption{The results of $\xi(\sigma,\pi)$ calculated directly from
$N$-body simulations of two different CDM models. Fig.~(a) shows the
average of the 9 available SCDM simulations, while Fig.~(b) shows the
average of the 5 available LCDM simulations. The different types of
contours are described in detail in the text of Section~\ref{xisp2sec}.
It is clear that the elongation of the contours in the line of sight
($\pi$) direction, caused by the non-linear galaxy velocity dispersion,
dominates the SCDM model. A similar effect is seen in the LCDM model,
although at a lower level.} \label{xispcdmfig}
\end{figure}

We have calculated the $\xi(\sigma,\pi)$ results from two different CDM
models (Efstathiou et al. 1985; Gazta\~{n}aga \& Baugh 1995; Eke et
al.  1996): standard CDM with $\Omega h = 0.5$, $b = 1.6$ (SCDM); and
CDM with $\Omega h = 0.2$, $b = 1$ and a cosmological constant
($\Lambda = 0.8$) to ensure a spatially flat cosmology (LCDM). In
Fig.~\ref{xispcdmfig}(a) we show the average $\xi(\sigma,\pi)$
calculated directly from the 9 SCDM $N$-body simulations which were
available to us, while Fig.~\ref{xispcdmfig}(b) shows the average from
the 5 available LCDM simulations. Given that these simulations are
fully volume limited with a well-defined mean density there are no
problems with estimator/weighting schemes and we estimate $\xi$ using
the methods described in Paper III for this case. For simplicity, we
also use the distant observer approximation (i.e. we imagine placing
the $N$-body cube at a large distance from the observer) and therefore
the line of sight direction can be assumed to be one specific
direction. We choose this to be the $z$-direction and hence $\sigma =
\sqrt{x^{2} + y^{2}}$, $\pi = z$ and the bins are cylindrical shells.

Fig.~\ref{xispcdmfig}(a) shows that the small scale velocity dispersion
in the SCDM model dominates the whole plot, elongating the contours
even on large scales ($\pi > 10h^{-1}$Mpc). The $\xi = 1$ contour cuts
the $\sigma$ axis between 4.5-5.0$h^{-1}$Mpc, which agrees well with
the real space correlation length of $\sim$5.0$h^{-1}$Mpc estimated in
Paper III. While there is a strong signal to be modelled for the
non-linear results, it is very doubtful that any useful information
will be obtained for the linear results of the SCDM model without a
more sophisticated approach than is attempted here.

Fig.~\ref{xispcdmfig}(b) shows that the small scale velocity dispersion
in the LCDM model is also the dominant feature on this plot, albeit
less pronounced than the SCDM model. Indeed, there is possible evidence
of a flattening in the $\pi$ direction on $\sim$20$h^{-1}$Mpc scales.
The $\xi = 1$ contour cuts the $\sigma$ axis between
6.0-6.5$h^{-1}$Mpc, which again agrees well with the real space
correlation length of $\sim$6.0$h^{-1}$Mpc estimated in Paper III.
Again, there is a strong signal for the non-linear results to model but
this time it appears possible that a sensible linear result could be
obtained from the LCDM model (see Section~6).

\subsection{Results from the Durham/UKST Survey} \label{durxispsec}

\begin{figure}
\centering
\centerline{\epsfxsize=8.5truecm \figinsert{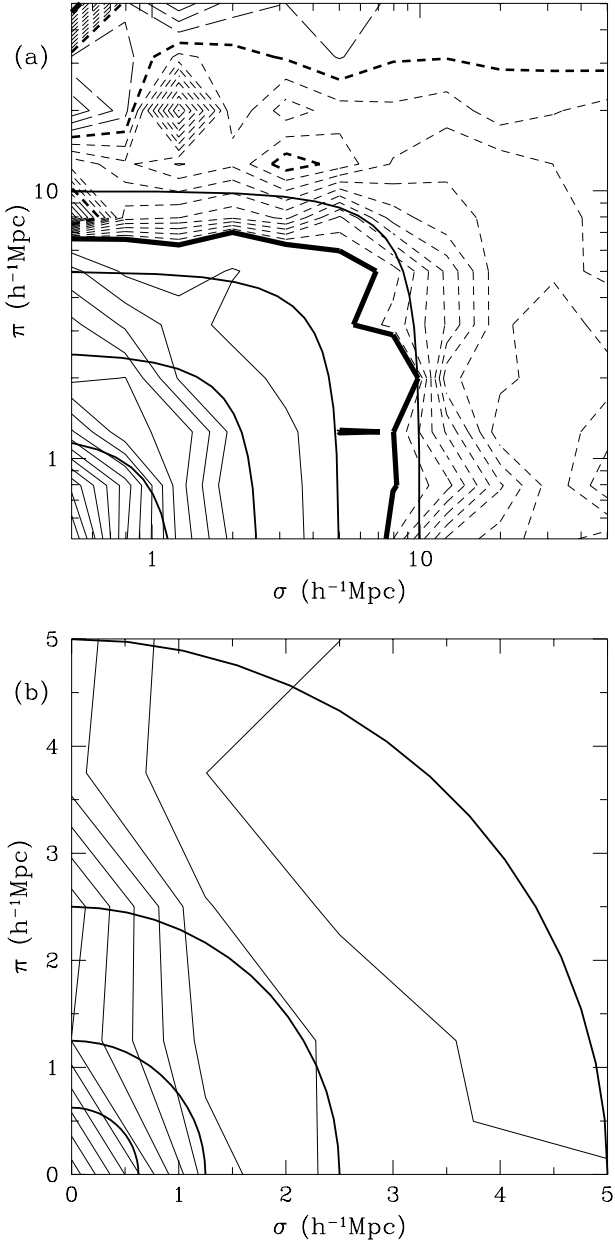}{0.0pt}}
\caption{The results of $\xi(\sigma,\pi)$ estimated from the
Durham/UKST survey using the optimal methods involving the estimator of
Hamilton (1993) and the weighting of Efstathiou (1988). Fig.~(a) plots
the results on a log-log scale to emphasize the large scale features,
while Fig.~(b) plots the results on a linear-linear scale to emphasize
the small scale features. The different types of contours are described
in detail in the text of Section~\ref{xisp2sec}. One immediately sees
that the noise levels in these diagrams are significantly higher than
those of the $N$-body simulations shown in Fig.~\ref{xispcdmfig}. These
contour plots also show that the non-linear velocity dispersion
elongates the contours in the line of sight ($\pi$) direction out to
5-10$h^{-1}$Mpc. However, on larger scales there is evidence for a
measurable flattening of the contours in the $\pi$ direction.}
\label{xispdurfig}
\end{figure}

In Fig.~\ref{xispdurfig} we show the results of $\xi(\sigma,\pi)$ from
the Durham/UKST survey as calculated using the optimal methods and
magnitude limits described in Ratcliffe et al. (1996b) and Paper III.
As mentioned in Section~\ref{xisp2sec} this includes the estimator of
Hamilton (1993) and the weighting of Efstathiou (1988). It is obvious
that the noise levels for the data are significantly higher than those
of the $N$-body simulations. Given the differences in signal between
the $N$-body simulations and the redshift survey this was to be
expected. As an aside, we note that the CDM mock catalogues used later
in this paper have similar noise levels to the data in plots like
these.  Fig.~\ref{xispdurfig} shows that the non-linear velocity
dispersion does not dominate the whole plot (unlike the CDM models of
Fig.~\ref{xispcdmfig}) and there is no significant elongation beyond
$\pi$$\sim$5-10$h^{-1}$Mpc. Also, on larger scales than these a visible
compression of the contours in the $\pi$ direction in seen. Therefore,
we are confident that believable non-linear and linear results can be
obtained from the Durham/UKST survey using the approach attempted
here.

\section{The Real Space 2-Point Correlation Function} \label{xirsec}
We define the projected 2-point correlation function, $w_{v}(\sigma)$,
by (e.g. Peebles 1980)
\begin{eqnarray}
w_{v}(\sigma) & = & \int_{-\infty}^{\infty} \xi(\sigma,\pi) d\pi ,\\
& = & 2 \int_{0}^{\infty} \xi(\sqrt{\sigma^{2} + \pi^{2}}) d\pi ,
\label{wvsigeqn}
\end{eqnarray}
where $\xi(\sqrt{\sigma^{2} + \pi^{2}})$ is the real space correlation
function. We calculate $w_{v}(\sigma)$ using
\begin{equation}
w_{v}(\sigma) = 2 \int_{0}^{\pi_{cut}} \xi(\sigma,\pi) d\pi ,
\label{wvsigeqn2}
\end{equation}
where the noise in $\xi(\sigma,\pi)$ at very large scales makes us
truncate the integral at some finite limit $\pi_{cut}$. In practice we
use a $\pi_{cut}$ of $30h^{-1}$Mpc and our results are insensitive to
raising this value.

\begin{figure}
\centering
\centerline{\epsfxsize=8.5truecm \figinsert{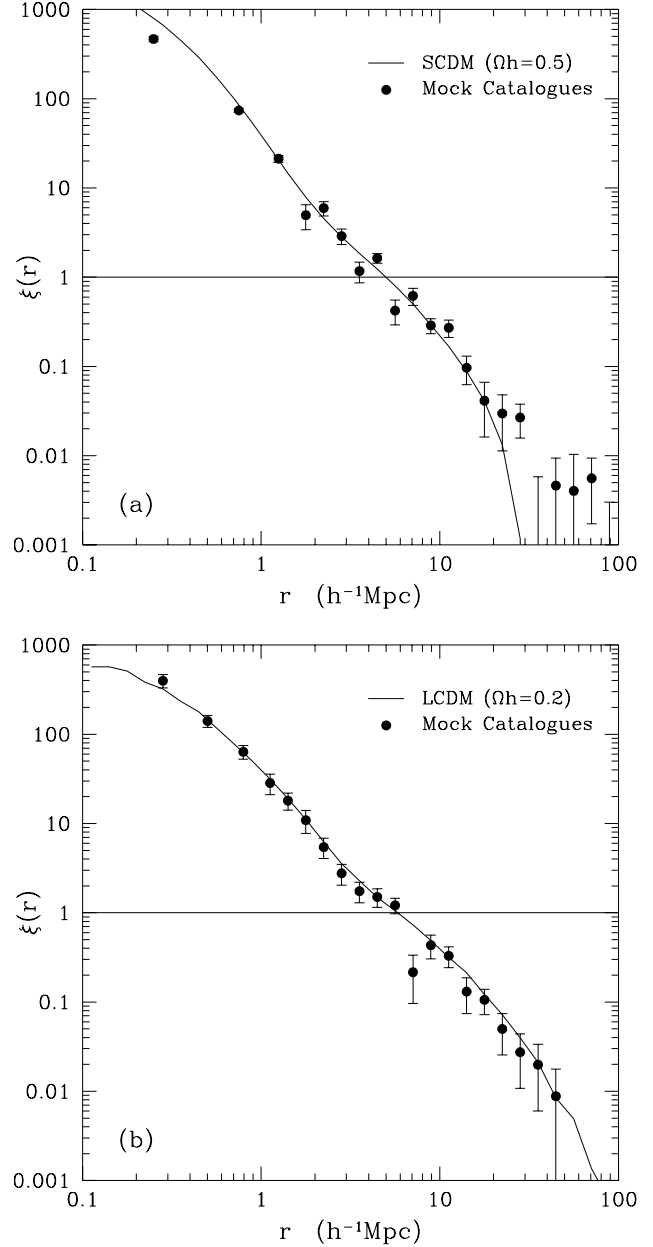}{0.0pt}}
\caption{Testing the methods of estimating $\xi(r)$ using the Abel
inversion of $w_{v}(\sigma)$ from (a) the SCDM and (b) the LCDM mock
catalogues. We show a comparison of the actual $\xi(r)$ calculated
directly from $N$-body simulations (solid line) with the average
$\xi(r)$ obtained from the mock catalogues (solid symbols). These
points came from taking each catalogue's optimally estimated
$\xi(\sigma,\pi)$, performing the integral in equation~\ref{wvsigeqn2}
to give a set of $w_{v}(\sigma)$'s, then carrying out the summations in
equation~\ref{xireqn} to give a set of $\xi(r)$'s and finally averaging
to obtain the results. The error bars show the $1\sigma$ standard
deviation on this mean assuming that each mock catalogue provides an
independent estimate of $\xi(r)$.} \label{xircdmfig}
\end{figure}

\clearpage

As we have discussed in Section~\ref{xispsec}, any catalogue which uses
redshifts to estimate distances suffers from the effects of galaxy
peculiar velocities. Therefore, only the redshift space 2-point
correlation function, $\xi(s)$, is directly observable from a redshift
survey. However, in terms of measuring and comparing the clustering
properties of the galaxy distribution with models of structure
formation, the object of fundamental interest is the real space 2-point
correlation function, $\xi(r)$. While $\xi(r)$ is not directly
observable from a redshift survey we will see that it is possible to
estimate it in an indirect manner. In this context we follow on from
results presented in Paper III.

\subsection{The Projected 2-Point Correlation Function}

Our methods of estimating $w_{v}(\sigma)$ were tested in Paper III
where the results from the Durham/UKST survey were first presented
(using the optimal methods mentioned in Section~\ref{durxispsec} for
calculating $\xi$). We then went on to model $w_{v}(\sigma)$ using a
power law for $\xi(r) = (r_{0}/r)^{\gamma}$ and estimated the real
space correlation length, $r_{0} = 5.1 \pm 0.3$, and slope, $\gamma =
1.6 \pm 0.1$. In this section, rather than assume a power law form, we
consider a direct inversion of equation~\ref{wvsigeqn} to give
$\xi(r)$.

\subsection{Abel Inversion} \label{abelsec}

Equation~\ref{wvsigeqn} can be mathematically inverted using the
generalized Abel equation to give (e.g. Lilje \& Efstathiou 1988)
\begin{equation}
\xi(r) = -\frac{1}{\pi} \int_{r}^{\infty}
\frac{d[w_{v}(\sigma)]}{d\sigma} \frac{d\sigma}{\sqrt{\sigma^{2} -
r^{2}}} .
\end{equation}
Saunders, Rowan-Robinson \& Lawrence (1992) consider the case when the
data is logarithmically binned and $w_{v}(\sigma)$ then takes the form
of a series of step functions, $w_{v}(\sigma_{i}) = w_{i}$, with
logarithmic spacing centered on $\sigma_{i}$. They then linearly
interpolate between each $w_{v}$ point to get around singularities in
the integral. In this case the $d[w_{v}(\sigma)]/d\sigma$ factor
simplifies to a constant value for each pair of $\sigma_{i}$ spacings.
The rest of the integral can be evaluated to give $\xi(r)$ at $r =
\sigma_{i}$
\begin{equation}
\xi(\sigma_{i}) = - \frac{1}{\pi} \sum_{j \geq i} \left[
\frac{w_{j+1}-w_{j}}{\sigma_{j+1}-\sigma_{j}} \right] \ln \left(
\frac{\sigma_{j+1} + \sqrt{\sigma_{j+1}^{2}-\sigma_{i}^{2}}}{\sigma_{j}
+ \sqrt{\sigma_{j}^{2}-\sigma_{i}^{2}}} \right) . \label{xireqn}
\end{equation}

We test this method using mock catalogues drawn from the SCDM and LCDM
$N$-body simulations. These mock catalogues have the same
angular/radial selection functions and completeness rates as the
Durham/UKST survey and their construction was described in detail in
Paper III. We selected the mock catalogues to sample independent
volumes of the simulations, giving a total of 18 and 15 to analyse from
the SCDM and LCDM simulations, respectively. The results of
$w_{v}(\sigma)$ from these two sets of mock catalogues were presented
in Paper III and Fig.~\ref{xircdmfig} shows the results of applying
equation~\ref{xireqn} to this $w_{v}(\sigma)$ data. The solid line
shows the actual $\xi(r)$ calculated directly from the $N$-body
simulations and therefore denotes the answer we are trying to obtain.
The solid symbols show the average $\xi(r)$ obtained from the mock
catalogues by taking each catalogue's optimally estimated
$\xi(\sigma,\pi)$, performing the integral in equation~\ref{wvsigeqn2}
to give a set of $w_{v}(\sigma)$'s, then carrying out the summations in
equation~\ref{xireqn} to give a set of $\xi(r)$'s and finally
averaging. The error bars show the $1\sigma$ standard deviation on this
mean assuming that each mock catalogue provides an independent estimate
of $\xi(r)$.

The SCDM results in Fig.~\ref{xircdmfig}(a) show that the mock
catalogues reproduce the actual $\xi(r)$ quite well out to
$\sim$25$h^{-1}$Mpc scales (with some degree of scatter about the
answer). On scales larger than this $\xi(r)$ is zero and therefore
there is no signal to invert, other than the noise in
$\xi(\sigma,\pi)$. The LCDM results in Fig.~\ref{xircdmfig}(b) show
that the mock catalogues reproduce the actual $\xi(r)$ very well out to
$\sim$40$h^{-1}$Mpc (apart from the point at $\sim$7$h^{-1}$Mpc which
is caused by a couple of negative $\xi(r)$ points produced in the
inversion process). Overall, we see that this method of inversion does
reproduce the actual $\xi(r)$ and its features quite well, albeit with
considerable scatter in places.

\begin{figure}
\centering
\centerline{\epsfxsize=8.5truecm \figinsert{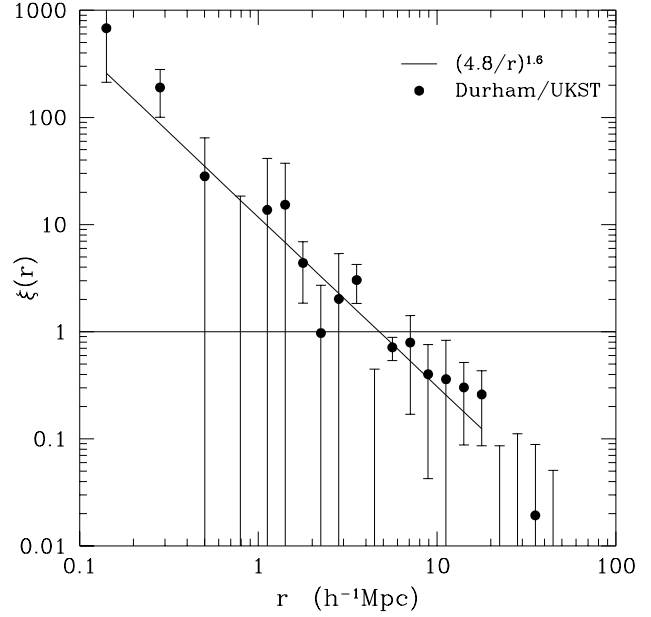}{0.0pt}}
\caption{Estimates of $\xi(r)$ from the Durham/UKST survey obtained by
Abel inversion of $w_{v}(\sigma)$. The points are calculated by taking
the optimally estimated $\xi(\sigma,\pi)$ from Fig.~\ref{xispdurfig},
evaluating the integral in equation~\ref{wvsigeqn2} to give
$w_{v}(\sigma)$ and then inverting this with equation~\ref{xireqn} to
produce $\xi(r)$. The solid line shows the best fitting power law to
the estimated $\xi(r)$ in the indicated range.} \label{xirabelfig}
\end{figure}

This method of estimating $\xi(r)$ is applied to the Durham/UKST survey
$\xi(\sigma,\pi)$ data of Fig.~\ref{xispdurfig}. The $w_{v}(\sigma)$
results were shown in Paper III and are used in equation~\ref{xireqn}
to produce the $\xi(r)$ shown in Fig.~\ref{xirabelfig}. The error bars
on this plot are obtained by splitting the Durham/UKST survey into 4
quadrants, applying the method to each quadrant and then estimating a
standard deviation of these results. These error bars are of a similar
size to those estimated on an individual LCDM mock catalogue. The solid
line shows the minimum $\chi^{2}$ fit of a power law $\xi(r) =
(r_{0}/r)^{\gamma}$ in the 0.1-20$h^{-1}$Mpc region. The best fit
parameters were $r_{0} = 4.8 \pm 0.5 h^{-1}$Mpc and $\gamma = 1.6 \pm
0.3$, which gave a $\chi^{2}$ of $\sim$8 for 15 degrees of freedom.
Note that the negative point at $\sim$4.5$h^{-1}$Mpc was removed from
this $\chi^{2}$ statistic to avoid biasing the fit. Errors on these
parameters come from the appropriate $\Delta\chi^{2}$ contour about the
minimum. However, given the correlated nature of these points we
anticipate that our quoted errors are more than likely an
underestimate. This should be adequate for the simple comparison done
here.

\subsection{Richardson-Lucy Inversion}

Richarson-Lucy inversion has recently been made popular in large-scale
structure applications by Baugh \& Efstathiou (1993) and here we apply
it to estimate $\xi(r)$ from $w_{v}(\sigma)$. Equation~\ref{wvsigeqn}
can be re-written as
\begin{equation}
w_{v}(\sigma) = \int_{\sigma}^{\infty} \xi(r)
\left(\frac{2r}{\sqrt{r^{2}-\sigma^{2}}}\right) dr , \label{wvsigeqn3}
\end{equation}
by changing the variable of integration from $\pi$ to $r$. If we then
define
\begin{eqnarray}
K(\sigma,r) & = & 0
\;\;\;\;\;\;\;\;\;\;\;\;\;\;\;\;\;\;\;\;\;\;\;\,\,\,\,\, {\rm for} \;\;\; 0
< r < \sigma , \\
& = & \frac{2r}{\sqrt{r^{2}-\sigma^{2}}} \;\;\;\;\;\;\;\;\;\;\;\,\,
{\rm for} \;\;\; \sigma < r < \infty ,
\end{eqnarray}
our integral becomes
\begin{equation}
w_{v}(\sigma) = \int_{0}^{\infty} \xi(r) K(\sigma,r) dr ,
\end{equation}
which is in the form of a Fredholm integral equation of the first
kind
\begin{equation}
\phi(x) = \int_{a}^{b} \psi(t) P(x|t) dt ,
\end{equation}
where $\phi(x)$ is the known (or observed) function, $\psi(t)$ is the
unknown function and $P(x|t)$ is the kernel of the integral equation.
Therefore, equation~\ref{wvsigeqn3} can be numerically inverted using
the method developed independently by Richardson (1972) and Lucy
(1974). Basically, one guesses an initial form for $\psi(t)$, predicts
$\phi(x)$ and uses this to update the original $\psi(t)$. This
iteration continues and after $n$ iterates
\begin{equation}
\phi^{n}(x) = \int_{a}^{b} \psi^{n}(t) P(x|t) dt , \label{RL1}
\end{equation}
with the next iterate of $\psi(t)$ given by
\begin{equation}
\psi^{n+1}(t) = \psi^{n}(t) \frac{\int_{a}^{b}
\frac{\tilde{\phi}(x)}{\phi^{n}(x)} P(x|t) dx}{\int_{a}^{b} P(x|t) dx}
, \label{RL2}
\end{equation}
where $\tilde{\phi}(x)$ is the actual observed function. To apply this
method to the logarithmically binned $w_{v}(\sigma)$ data the integrals
in equations~\ref{RL1} and~\ref{RL2} are approximated by the following
summations
\begin{eqnarray}
w_{v}^{n}(\sigma_{j}) & = & \sum_{i=1}^{N} \xi^{n}(r_{i})
K(\sigma_{j},r_{i}) r_{i} \Delta \ln r , \label{xireal2a} \\
\xi^{n+1}(r_{i}) & = & \xi^{n}(r_{i}) \frac{\sum_{j=1}^{M}
\frac{\tilde{w}_{v}(\sigma_{j})}{w_{v}^{n}(\sigma_{j})}
K(\sigma_{j},r_{i}) \sigma_{j} \Delta \ln \sigma}{\sum_{j=1}^{M}
K(\sigma_{j},r_{i}) \sigma_{j} \Delta \ln \sigma} , \label{xireal2b}
\end{eqnarray}
where $M$ is the number of $w_{v}(\sigma)$ bins and $N=M/2$ is the
number of $\xi(r)$ bins. Our spacing in $\sigma$ is
$\Delta\lg\sigma=0.1$ and hence our spacing in $r$ is $\Delta\lg
r=0.2$. Obviously one cannot get back more data points than are put in
and $N \leq M$. In general, our choice of $N=M/2$ should assure a
fairly smooth answer.

There are two points worth noting about Richardson-Lucy algorithms.
Firstly, there is no constraint on how many iterations are required for
convergence to a stable answer. Therefore, there is no specific rule to
know when to stop iterating. Too many iterations will cause convergence
to the small scale noise features in the data, while too few implies
that the results have not yet converged to the larger scale signal.
Experience with Richardson-Lucy techniques shows that $\sim$10
iterations are generally required (e.g. Lucy 1994). Secondly, this
method assumes that the function $\psi(t) \geq 0$. This is not always
the case for our function $\xi(r)$. However, this is not too worrying
as $\xi(r)$ is only likely to go negative on large scales when it is
nearly zero, this is where our inversion process will be least
believable anyway. Also, other authors (e.g. Baugh \& Efstathiou 1993;
Baugh 1996) have applied similar inversion techniques to the angular
correlation function to estimate the power spectrum (always positive)
and real space correlation function (negative tail) and find very
consistent results.

\begin{figure}
\centering
\centerline{\epsfxsize=8.5truecm \figinsert{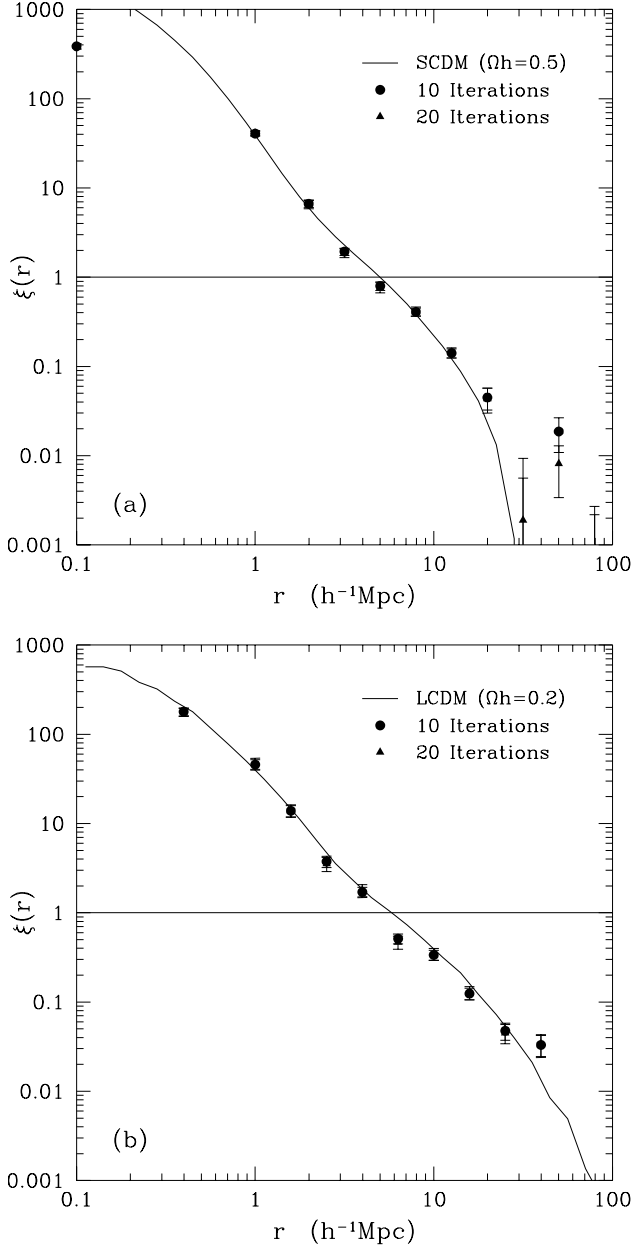}{0.0pt}}
\caption{Testing the methods of estimating $\xi(r)$ using
Richardson-Lucy inversion of $w_{v}(\sigma)$ from (a) the SCDM and (b)
the LCDM mock catalogues. We show a comparison of the actual $\xi(r)$
calculated directly from $N$-body simulations (solid line) with the
average $\xi(r)$ obtained from the mock catalogues (solid symbols).
These points came from taking each catalogue's optimally estimated
$\xi(\sigma,\pi)$, performing the integral in equation~\ref{wvsigeqn2}
to give a set of $w_{v}(\sigma)$'s, then carrying out the iteration
between equations~\ref{xireal2a} and~\ref{xireal2b} to give a set of
$\xi(r)$'s and finally averaging to obtain the results. The circles
show the results on stopping after 10 iterations, while the triangles
show the same results after 20 iterations. The error bars show the
$1\sigma$ standard deviation on this mean assuming that each mock
catalogue provides an independent estimate of $\xi(r)$.}
\label{xircdmfig2}
\end{figure}

We again test these methods using the SCDM and LCDM mock catalogues.
Fig.~\ref{xircdmfig2} shows the results of applying
equations~\ref{xireal2a} and~\ref{xireal2b} to the $w_{v}(\sigma)$
data. Again, the solid line shows the actual $\xi(r)$ calculated
directly from the $N$-body simulations and the solid symbols show the
$\xi(r)$ obtained from the mock catalogues by individually inverting
$w_{v}(\sigma)$ with the iteration process and averaging at the end.
The circles show the results on stopping after 10 iterations, while the
triangles show the results after 20 iterations. The results were found
to be independent of the initial form of $\xi(r)$, therefore a simple
power law was used. The error bars show the $1\sigma$ standard
deviation on this mean assuming that each mock catalogue provides an
independent estimate of $\xi(r)$.

The SCDM results in Fig.~\ref{xircdmfig2}(a) show that the mock
catalogues reproduce the actual $\xi(r)$ very well out to
$\sim$30$h^{-1}$Mpc scales. On larger scales noise in the data
dominates and this is the limit where we believe our inversion
technique. It appears that the results do not change very much with
further iteration. The LCDM results in Fig.~\ref{xircdmfig2}(b) show a
similar form to those of the SCDM mock catalogues and reproduce the
actual $\xi(r)$ very well out to $\sim$40$h^{-1}$Mpc scales. Again, on
larger scales noise in the data dominates and this is our believable
inversion limit. Further iteration does not damage the stability of the
inversion although the results do become less smooth. We note that this
inversion technique does not appear to do well at the end-points of the
integration range, this is an artifact of the finite integration range
(which assumes that the signal is zero outside of this range) and the
large bins used in the summations. Therefore, our results could be
improved by interpolating between the $w_{v}(\sigma)$ points and then
inverting this function given that we could calculate it at many
points. Given the uncertainties involved in $\xi(\sigma,\pi)$ this is
not attempted here. Overall, this method of inversion reproduces the
actual $\xi(r)$ and its features very well and is arguably smoother
than the Abel inversion. However, this is offset by only having an
estimate at half the number of $r$ points.

\begin{figure}
\centering
\centerline{\epsfxsize=8.5truecm \figinsert{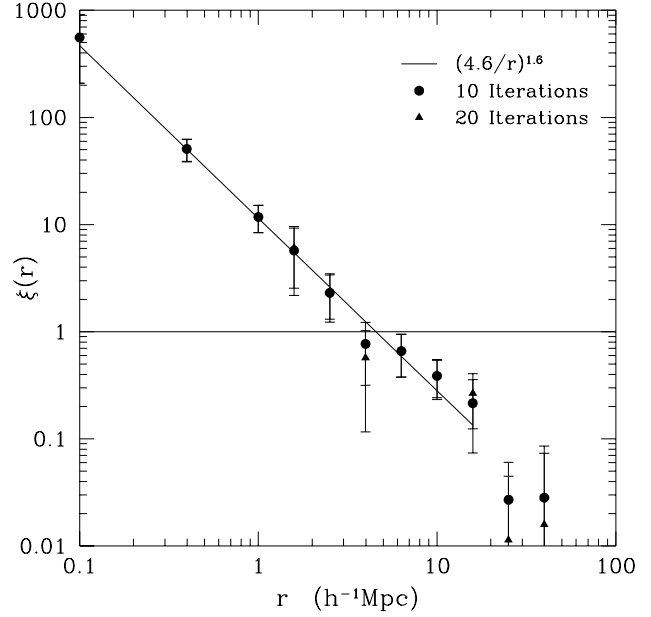}{0.0pt}}
\caption{Estimates of $\xi(r)$ from the Durham/UKST survey obtained by
Richardson-Lucy inversion of $w_{v}(\sigma)$. The solid points are
calculated by taking the optimally estimated $\xi(\sigma,\pi)$ from
Fig.~\ref{xispdurfig}, evaluating the integral in
equation~\ref{wvsigeqn2} to give $w_{v}(\sigma)$ and then inverting
using the iteration process involving equations~\ref{xireal2a}
and~\ref{xireal2b} to produce $\xi(r)$. The circles show the results on
stopping after 10 iterations, while the triangles show the results
after 20 iterations. The solid line shows the best fitting power law to
the estimated $\xi(r)$ in the indicated range.} \label{xirrlfig}
\end{figure}

This method of estimating $\xi(r)$ is applied to the Durham/UKST survey
$\xi(\sigma,\pi)$ data of Fig.~\ref{xispdurfig}. The $w_{v}(\sigma)$
results were shown in Paper III and are used in
equations~\ref{xireal2a} and~\ref{xireal2b} to produce the $\xi(r)$
shown in Fig.~\ref{xirrlfig}. The error bars on this plot are obtained
by splitting the Durham/UKST survey into 4 quadrants, applying the
method to each quadrant and then estimating a standard deviation of
these results. These error bars are of a similar size to those
estimated on an individual LCDM mock catalogue. It can be seen that
further iteration does not improve the results, in fact it makes them
slightly less smooth. The solid line shows the minimum $\chi^{2}$ fit
of a power law $\xi(r) = (r_{0}/r)^{\gamma}$ in the 0.1-20$h^{-1}$Mpc
region. The best fit parameters were $r_{0} = 4.6 \pm 0.6 h^{-1}$Mpc
and $\gamma = 1.6 \pm 0.1$, which gave a $\chi^{2}$ of $\sim$2 for 8
degrees of freedom. Note that the point at $\sim$4.5$h^{-1}$Mpc (which
was negative for the Abel inversion) is still low. However, it does not
bias the fit because the larger binning in this method has essentially
averaged over it. Errors on these parameters come from the appropriate
$\Delta\chi^{2}$ contour about the minimum. We make a similar comment
to before about the correlated nature of these points implying an
underestimated error on these parameters. Again, this should be
adequate for the simple comparison done here.

\subsection{Discussion} \label{realreddissec}

Before discussing the results in detail we make a few comments about
the inversion methods themselves. Firstly, given that the Abel
inversion is a point by point process, the noise in $w_{v}(\sigma)$ is
also inverted. This is similar to what is seen if the Richardson-Lucy
inversion is allow to continue iterating, namely it converges to the
small scale noise features in the data. Therefore, the smoothest answer
is produced by the Richardson-Lucy inversion after $\sim$10 iterations.
Secondly, both of these methods do not do particularly well at the end
points of the integration range. On the large scales this was to be
expected as it is here that the signal is almost zero and the noise
dominates. On small scales this is due to a combination of the cutoff
at 0.1$h^{-1}$Mpc and the approximation of the integrals in
equations~\ref{xireal2a} and~\ref{xireal2b} by finite summations with
quite large bins. Thirdly, as a consistency check on the
Richardson-Lucy method, we have examined the $w_{v}(\sigma)$ produced
by the iteration procedure. We find that this $w_{v}(\sigma)$ is in
excellent agreement with the original one obtained directly from the
data.  Finally, while we have only shown the results of testing these
inversion methods on the CDM mock catalogues, we also tested these
techniques on other data sets. We constructed simple power laws in
$\xi(\sigma,\pi)$ and also more complicated power laws with break
features, both with and without noise. The methods passed all of these
tests and therefore we are confident of applying them to the
Durham/UKST survey to give believable results for $\xi(r)$.

We now discuss the $\xi(r)$ results obtained from the Durham/UKST
survey. Figs.~\ref{xirabelfig} and~\ref{xirrlfig} show that the results
from these two different inversion techniques agree well. The Abel
method is generally noisier then the Richardson-Lucy method but has
$\xi(r)$ estimates at twice as many points. Comparing the power law
fits gives very good agreement in both correlation length and slope:
$r_{0} = 4.8 \pm 0.5 h^{-1}$Mpc and $\gamma = 1.6 \pm 0.3$ for the Abel
method, and $r_{0} = 4.6 \pm 0.6 h^{-1}$Mpc and $\gamma = 1.6 \pm 0.1$
for the Richardson-Lucy method. These values are also in good agreement
with those estimated by modelling $\xi(r)$ as a power law and fitting
this model to $w_{v}(\sigma)$, namely $r_{0} = 5.1 \pm 0.3$ and slope
$\gamma = 1.6 \pm 0.1$ (Paper III). We will use these estimates of
$\xi(r)$ from the Durham/UKST survey in Section~\ref{linsec} when
comparing the real and redshift space correlation functions to estimate
a value of $\Omega^{0.6}/b$.

\begin{figure}
\centering
\centerline{\epsfxsize=8.5truecm \figinsert{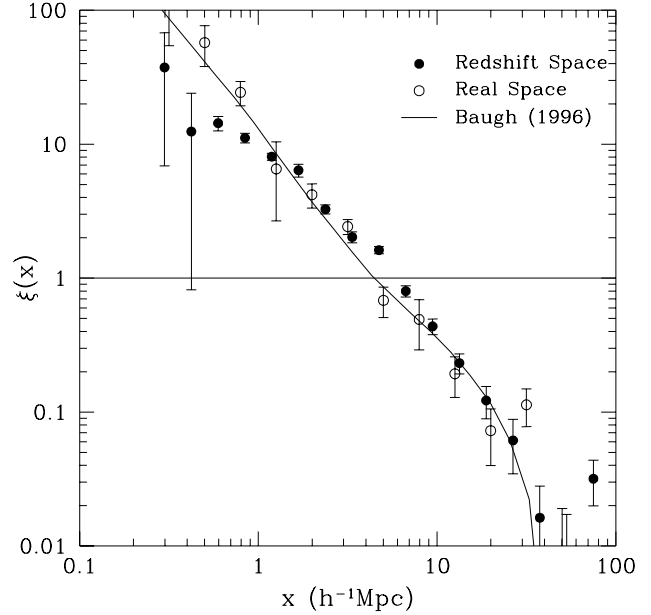}{0.0pt}}
\caption{Comparison of the real and redshift space 2-point correlation
functions obtained by combining the data from the Durham/UKST survey
with that of the APM-Stromlo survey and the Las Campanas survey
(Loveday et al. 1992; Loveday et al. 1995; Lin et al. 1996; Tucker et
al. 1996). The solid circles show the redshift space $\xi(s)$, which
is an error weighted mean of the 3 surveys. The open circles show the
real space $\xi(r)$ obtained from an error weighted combination of the
3 different Abel inverted $w_{v}(\sigma)$'s. All of the error bars
assume that each survey provides a statistically independent estimate
of $\xi$. The solid line shows Baugh's (1996) real space $\xi(r)$
estimate which comes from the inversion of the APM $w(\theta)$.}
\label{xirallfig}
\end{figure}

To produce a final estimate of the real and redshift space 2-point
correlation functions we have combined the data from the Durham/UKST
survey with that of the APM-Stromlo survey and the Las Campanas survey
(Loveday et al. 1992; Loveday et al. 1995; Lin et al. 1996; Tucker et
al. 1996). These results are shown in Fig.~\ref{xirallfig}. For the
redshift space 2-point correlation function (solid circles) we have
combined the different data sets directly using an error weighted mean
of the 3 surveys. For the real space 2-point correlation function (open
circles) we use the Abel inversion method (for ease) on the 3 different
$w_{v}(\sigma)$'s and then combine in an error weighted manner. All of
the combined estimates use the weighting of Efstathiou (1988) and the
estimator of Hamilton (1993). The error bars assume that each survey
provides a statistically independent estimate of $\xi$. For comparison,
we plot the real space 2-point correlation function (solid line)
estimated from the inversion of the APM $w(\theta)$ by Baugh (1996).

We see that the real space $\xi(r)$ appears to quite nicely approximate
a featureless single power law out to $\sim$20$h^{-1}$Mpc. The high
point at $\sim$30$h^{-1}$Mpc is at the end of the inversion region,
where this technique becomes unreliable. This estimate agrees well with
that of Baugh (1996), who also sees a slight `shoulder' feature on
$5-25h^{-1}$Mpc scales. The redshift space $\xi(s)$ appears to have a
change of shape in both amplitude and slope near $\sim$5$h^{-1}$Mpc.
The smaller scale effect is probably due to the velocity dispersion,
which smooths the signal out, while the larger scale effect is probably
due to the linear infall of galaxies, which amplifies the signal. These
effects are discussed in more detail in Sections~\ref{nonlinsec}
and~\ref{linsec}. Finally, we make a comment about the `shoulder'
feature seen in Baugh's (1996) $\xi(r)$. First evidence for such a
feature was observed in $\xi(s)$ by Shanks et al. (1983) and Shanks et
al. (1989) who noted the systematic increase between the amplitude of
$\xi(s)$ at $r \simeq 7h^{-1}$Mpc and the lower amplitudes obtained from
$w_{v}(\sigma)$ at smaller scales, as continue to be found here (see
Fig.~7). Some of this excess in $\xi(s)$ is now thought to be due to
redshift-space distortion from infall (see Section~6) with some
contribution from a real-space `shoulder' in $\xi(r)$, although the
evidence is stronger in the 2-D data of Baugh (1996) than in the
redshift survey data presented here.

\section{Non-Linear Scales: The One-Dimensional Pairwise Velocity
Dispersion} \label{nonlinsec}

We now use standard modelling techniques to estimate the
one-dimensional rms galaxy pairwise velocity dispersion, $\veldisp$,
from the Durham/UKST survey. As we saw in Section~\ref{xispsec} this
velocity dispersion visibly elongates the $\xi(\sigma,\pi)$ contours in
the line of sight direction and this is now quantitatively measured.

\subsection{Modelling $\xi(\sigma,\pi)$} \label{xispmodelsec}

We follow the modelling of Peebles (1980) and define ${\bf v}$ to be
the peculiar velocity of a galaxy above the Hubble flow, therefore
${\bf w} = {\bf v}_{i} - {\bf v}_{j}$ is the peculiar velocity
difference of two galaxies separated by a vector ${\bf r}$. Let $g({\bf
r},{\bf w})$ be the distribution function of ${\bf w}$. Therefore
\begin{equation}
1 + \xi(\sigma,\pi) = \int \left[ 1 + \xi(r) \right] g({\bf r},{\bf w})
d^{3}w , \label{xispveldisp}
\end{equation}
where
\begin{equation}
r^{2} = \sigma^{2} + r_{z}^{2} \;\;\;\; , \;\;\;\; r_{z} = \pi -
\frac{w_{z}}{H_{0}} ,
\end{equation}
and $w_{z}$ is the component of ${\bf w}$ parallel to the line of
sight, which for simplicity is called the $z$ direction. It is common
to assume that $g$ is a slowly varing function of ${\bf r}$, such that
$g({\bf r},{\bf w}) = g({\bf w})$. Therefore, one can make the
approximation
\begin{equation}
\int dw_{x} \int dw_{y} \; g({\bf w}) = f(w_{z}) . \label{intgtof}
\end{equation}
Equation~\ref{xispveldisp} then becomes
\begin{equation}
1 + \xi(\sigma,\pi) = \int \left[ 1 + \xi(r) \right] f(w_{z}) dw_{z} ,
\end{equation}
which further reduces to
\begin{equation}
\xi(\sigma,\pi) = \int_{-\infty}^{\infty} \xi(r) f(w_{z}) dw_{z} ,
\end{equation}
when the unit normalisation of $f(w_{z})$ is considered. A streaming
model which describes the relative bulk motion of galaxies towards (or
away from) each other can be incorporated as follows
\begin{equation}
g({\bf r},{\bf w}) = g({\bf w} - {\bf \hat{r}}v(r)) ,
\end{equation}
where $v(r)$ is the streaming model in question.
Equation~\ref{xispveldisp} then becomes
\begin{eqnarray}
& & 1 + \xi(\sigma,\pi) = \int \left[ 1 + \xi(r) \right] f(w_{z} -
v(r_{z})) dw_{z} ,\\
& = & \int_{-\infty}^{\infty} \left[ 1 + \xi\left( \sqrt{\sigma^{2} +
r_{z}^{2}} \right) \right] f\left[ w_{z} - v(r_{z}) \right] dw_{z} .
\end{eqnarray}
Obviously, we require models for the real space 2-point correlation
function, $\xi(r)$, the distribution function, $f(w_{z})$, and the
streaming motion, $v(r_{z})$. The real space 2-point correlation
function is simply modelled by a power law, $\xi(r) =
(r_{0}/r)^{\gamma}$, which should be accurate out to
$\sim$20$h^{-1}$Mpc. For the distribution function we tried both
exponential and Gaussian functions and found that the exponential
provided a significantly better fit to the shape of the $N$-body
simulation results and so is used exclusively here
\begin{equation}
f(w_{z}) = \frac{1}{\sqrt{2} \veldisp} \exp \left[ -\sqrt{2}
\frac{\left| w_{z} \right|}{\veldisp} \right] ,
\end{equation}
where $\veldisp$ is the rms pairwise velocity dispersion, namely the
second moment of the distribution function $f(w_{z})$
\begin{equation}
< w^{2} > = \int_{-\infty}^{\infty} f(w_{z}) w_{z}^{2} dw_{z} .
\end{equation}
A realistic streaming model might be expected to depend on the
clustering, biasing and mean mass density of the universe. The infall
model of Bean et al. (1983) takes the maximal approach by assuming
$\Omega = 1$, $b = 1$ and uses the second BBGKY equation (e.g. Peebles
1980) to give
\begin{equation}
v(r_{z}) = - H_{0} r_{z} \left[\frac{\xi(r_{z})}{1 + \xi(r_{z})}
\right] . \label{streammod}
\end{equation}
We favour this streaming model in the analysis presented here.

\subsection{Testing the Modelling}

We have tested this modelling with both the CDM simulations and mock
catalogues. Before describing these tests we first present the answers
we are trying to reproduce. We have calculated the values of $\veldisp$
directly from the simulations and find that $\veldisp \simeq 950$ and
$750$ kms$^{-1}$ on $\sim$1$h^{-1}$Mpc scales for the SCDM and LCDM
models, respectively. In Paper II we estimated the real space $\xi(r)$
correlation lengths to be $\sim$5.0$h^{-1}$Mpc and $\sim$6.0$h^{-1}$Mpc
for the SCDM and LCDM models, respectively. For both of these models
the slope of the real space $\xi(r)$ was $\sim$2.2.

\begin{figure*}
\centering
\centerline{\epsfxsize=17.0truecm \figinsert{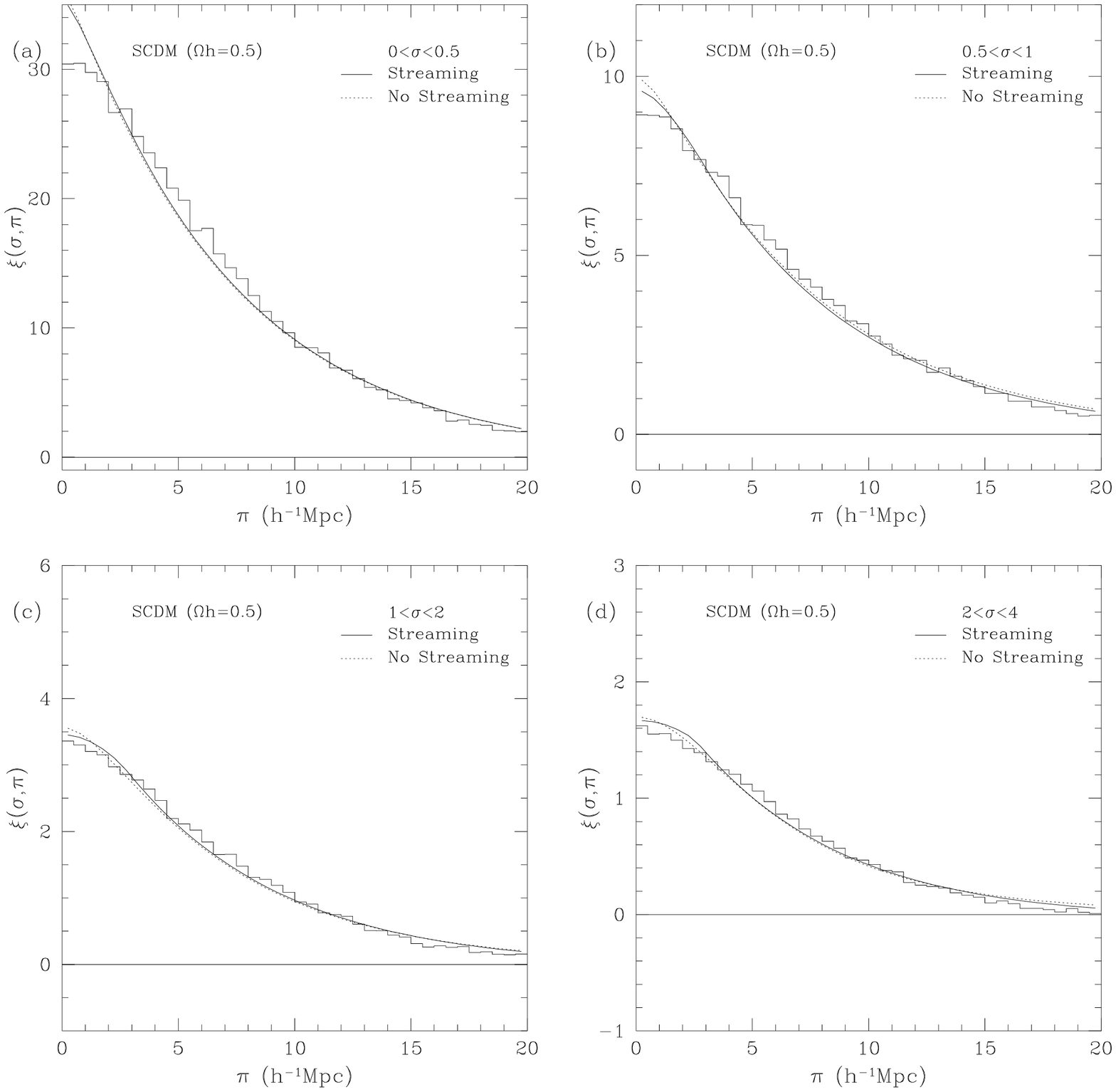}{0.0pt}}
\caption{Histograms of $\xi(\sigma,\pi)$ estimated from the SCDM
$N$-body simulations as a function of $\pi$ for different (constant)
values of $\sigma$. The solid curve shows the minimum $\chi^{2}$ fit
using the modelling of Section~\ref{xispmodelsec} with a streaming
model, while the dotted curve shows the fit without the streaming
model.} \label{xipiscdmfig}
\end{figure*}

\begin{figure*}
\centering
\centerline{\epsfxsize=17.0truecm \figinsert{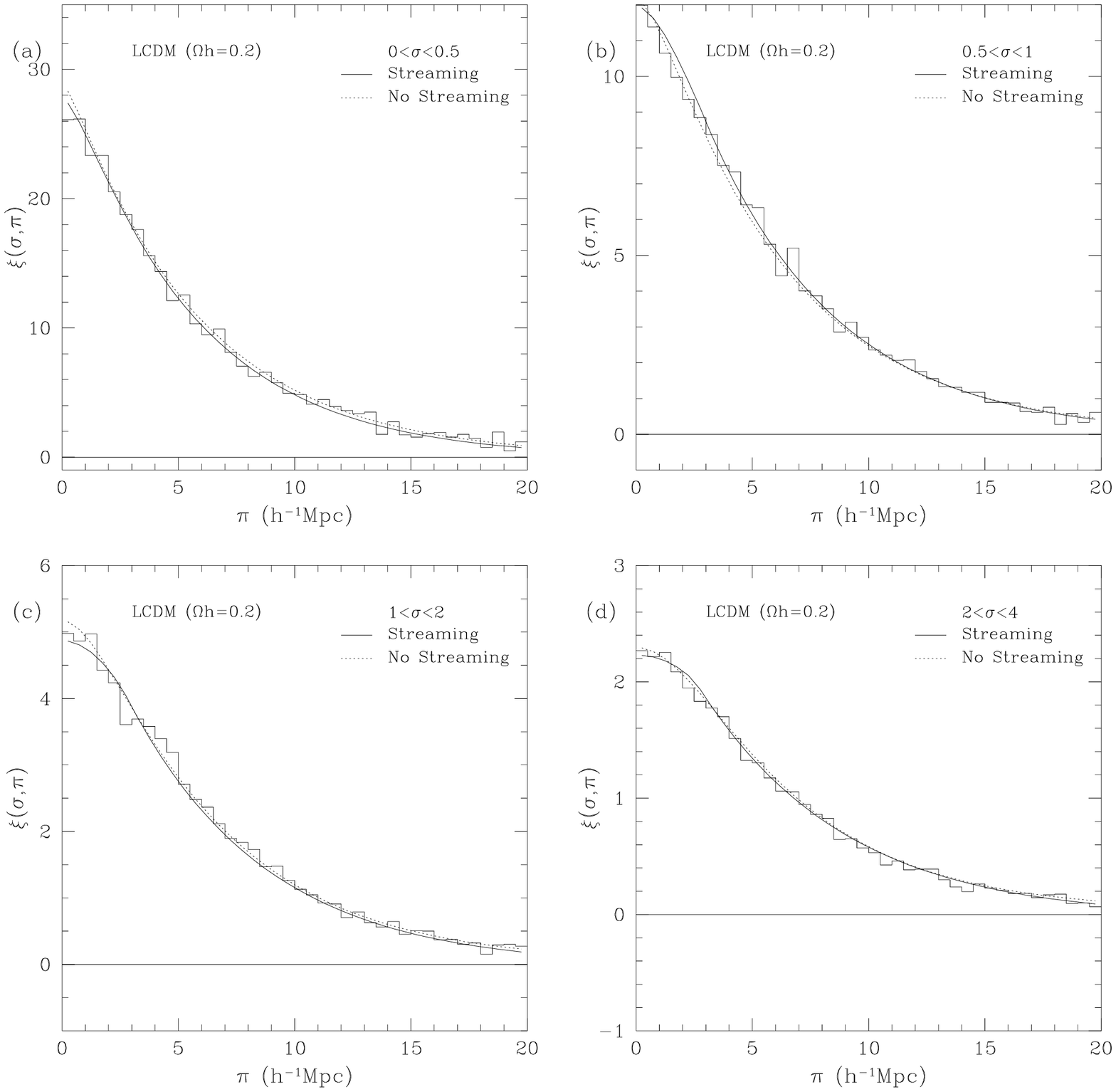}{0.0pt}}
\caption{Histograms of $\xi(\sigma,\pi)$ estimated from the LCDM
$N$-body simulations as a function of $\pi$ for different (constant)
values of $\sigma$. The solid curve shows the minimum $\chi^{2}$ fit
using the modelling of Section~\ref{xispmodelsec} with a streaming
model, while the dotted curve shows the fit without the streaming
model.} \label{xipilcdmfig}
\end{figure*}

In the fitting procedure there are three parameters which can be
estimated, $\veldisp$, $r_{0}$ and $\gamma$. Given our available
computing constraints we only choose to fit for two of these and fix
$\gamma$ to be a constant value. The results obtained from this
procedure are insensitive to the value of $\gamma$ chosen, provided a
realistic value is used. For the CDM simulations we use $\gamma =
2.2$.  Also, when the streaming model is used we assume that $r_{0} =$
5.0-6.0$h^{-1}$Mpc in equation~\ref{streammod} only, again the fits are
relatively insensitive to the value chosen. We fit $\veldisp$ and
$r_{0}$ using an approximate $\chi^{2}$ statistic in the range
0-20$h^{-1}$Mpc for four different values of $\sigma$, with the
standard deviations from the simulations or mock catalogues being used
accordingly in the $\chi^{2}$ statistic. The results of these fits
using the $\xi(\sigma,\pi)$ estimated directly from the $N$-body
simulations are shown in Figs.~\ref{xipiscdmfig} and~\ref{xipilcdmfig}
for the SCDM and LCDM models, respectively, with the best fit values
given in Table~\ref{xipicdmtab}. The solid histogram denotes the
averaged $\xi(\sigma,\pi)$ in the quoted $\sigma$ range, while the
solid and dotted lines show the fits with and without the streaming
model, respectively. Similarly, Table~\ref{xipicdm2tab} shows the
results of the fits to the optimally estimated $\xi(\sigma,\pi)$ from
the SCDM and LCDM mock catalogues. The quoted error bars on the mock
catalogue results come from the $1\sigma$ standard deviation between
the mock catalogues and therefore reflects the error on an individual
mock catalogue.

\begin{table}
\begin{center}
\caption{Minimum $\chi^{2}$ fits for $\veldisp$ and $r_{0}$ from the
CDM $N$-body simulations using the exponential distribution function
with and without the streaming model.} \label{xipicdmtab}
\begin{tabular}{cccc}
{\large $\sigma$} & {\large $\veldisp$} & {\large $r_{0}$} & {\large
$\chi^{2}$} \\
{\large ($h^{-1}$Mpc)} & {\large (kms$^{-1}$)} & {\large ($h^{-1}$Mpc)}
& {\large (N$_{bin} = 40$)} \\
& & & \\
\multicolumn{4}{c}{{\large SCDM}} \\
\multicolumn{4}{c}{{\large Streaming Model}} \\
$[0,0.5]$ & 990 & 5.1 & 40.0 \\
$[0.5,1]$ & 1000 & 5.3 & 24.7 \\
$[1,2]$ & 980 & 4.8 & 21.2 \\
$[2,4]$ & 950 & 4.8 & 34.4 \\
\multicolumn{4}{c}{{\large No Streaming Model}} \\
$[0,0.5]$ & 970 & 5.1 & 46.3 \\
$[0.5,1]$ & 960 & 5.4 & 32.3 \\
$[1,2]$ & 800 & 4.9 & 28.7 \\
$[2,4]$ & 550 & 5.1 & 48.2 \\
& & & \\
\multicolumn{4}{c}{{\large LCDM}} \\
\multicolumn{4}{c}{{\large Streaming Model}} \\
$[0,0.5]$ & 770 & 4.1 & 13.8 \\
$[0.5,1]$ & 810 & 5.3 & 7.8 \\
$[1,2]$ & 850 & 5.4 & 16.5 \\
$[2,4]$ & 910 & 5.6 & 11.1 \\
\multicolumn{4}{c}{{\large No Streaming Model}} \\
$[0,0.5]$ & 780 & 4.2 & 14.2 \\
$[0.5,1]$ & 760 & 5.4 & 8.1 \\
$[1,2]$ & 720 & 5.6 & 14.0 \\
$[2,4]$ & 570 & 5.9 & 9.0 \\
\end{tabular}
\end{center}
\end{table}

\begin{table}
\begin{center}
\caption{Minimum $\chi^{2}$ fits for $\veldisp$ and $r_{0}$ from the
CDM mock catalogues using the exponential distribution function with
and without the streaming model.} \label{xipicdm2tab}
\begin{tabular}{ccc}
{\large $\sigma$} & {\large $\veldisp$} & {\large $r_{0}$} \\
{\large ($h^{-1}$Mpc)} & {\large (kms$^{-1}$)} & {\large ($h^{-1}$Mpc)} \\
& & \\
\multicolumn{3}{c}{{\large SCDM}} \\
\multicolumn{3}{c}{{\large Streaming Model}} \\
$[0,0.5]$ & $930 \pm 190$ & $5.0 \pm 0.4$ \\
$[0.5,1]$ & $970 \pm 180$ & $5.3 \pm 0.4$ \\
$[1,2]$ & $920 \pm 260$ & $4.6 \pm 0.5$ \\
$[2,4]$ & $810 \pm 180$ & $4.5 \pm 0.6$ \\
\multicolumn{3}{c}{{\large No Streaming Model}} \\
$[0,0.5]$ & $920 \pm 190$ & $5.0 \pm 0.4$ \\
$[0.5,1]$ & $910 \pm 170$ & $5.3 \pm 0.4$ \\
$[1,2]$ & $750 \pm 230$ & $4.8 \pm 0.4$ \\
$[2,4]$ & $450 \pm 130$ & $5.0 \pm 0.4$ \\
& & \\
\multicolumn{3}{c}{{\large LCDM}} \\
\multicolumn{3}{c}{{\large Streaming Model}} \\
$[0,0.5]$ & $610 \pm 180$ & $3.8 \pm 0.3$ \\
$[0.5,1]$ & $790 \pm 230$ & $5.2 \pm 0.7$ \\
$[1,2]$ & $840 \pm 130$ & $5.2 \pm 0.7$ \\
$[2,4]$ & $780 \pm 150$ & $5.2 \pm 0.7$ \\
\multicolumn{3}{c}{{\large No Streaming Model}} \\
$[0,0.5]$ & $590 \pm 170$ & $3.9 \pm 0.3$ \\
$[0.5,1]$ & $750 \pm 220$ & $5.2 \pm 0.6$ \\
$[1,2]$ & $690 \pm 130$ & $5.3 \pm 0.7$ \\
$[2,4]$ & $440 \pm 140$ & $5.6 \pm 0.6$ \\
\end{tabular}
\end{center}
\end{table}

The results from the CDM simulations show that the streaming model only
becomes important (in terms of producing consistent results for
$\veldisp$) when \mbox{$\sigma >$ 1-2$h^{-1}$Mpc}. Of course, this
assumes that $\veldisp$ does not vary with $\sigma$. The best fit
values to the CDM simulation results (with streaming), had $\veldisp =
980 \pm 22$ kms$^{-1}$, $r_{0} = 5.00 \pm 0.24 h^{-1}$Mpc for the SCDM
model and $\veldisp = 835 \pm 60$ kms$^{-1}$, $r_{0} = 5.12 \pm 0.69
h^{-1}$Mpc for the LCDM model. These quoted errors simply come from the
scatter in the best fit values of Table~\ref{xipicdmtab}. The values of
$\veldisp$ can be compared with those estimated directly from the
$N$-body simulations on $\sim$1$h^{-1}$Mpc scales, namely $950$ and
$750$ kms$^{-1}$ for the SCDM and LCDM models, respectively. This
agreement is adequate given the slightly \mbox{ad hoc} assumption of an
exponential distribution function. The values of $r_{0}$ can be
compared with the approximate real space values estimated in Paper III,
namely $5.0h^{-1}$Mpc and $6.0h^{-1}$Mpc for the SCDM and LCDM models,
respectively. Again, the agreement is adequate in both cases although
slightly small for the LCDM model. However, closer inspection of the
LCDM $\xi(r)$ plot on $r \leq 1h^{-1}$Mpc scales shows that $r_{0}$ is
slightly lower in this region, $\sim$5.0$h^{-1}$Mpc, which probably
explains our measurement.

The results from the CDM mock catalogues confirm the conclusions drawn
from the CDM simulation results. All of the $\veldisp$ values in
Table~\ref{xipicdm2tab} agree well with their counterparts in
Table~\ref{xipicdmtab} given the quoted errors (on an individual mock
catalogue). A similar statement can be made for the values of $r_{0}$
estimated from these mock catalogues.

Overall, our tests confirm that the correct $\veldisp$ and $r_{0}$ can
be reproduced by this method which models the elongation in
$\xi(\sigma,\pi)$. This holds for both the $\xi(\sigma,\pi)$ estimated
directly from the $N$-body simulations and that estimated using the
optimal techniques for $\xi$ from the mock catalogues, albeit with
larger scatter in this case.

\subsection{Results from the Durham/UKST Survey}

\begin{figure*}
\centering
\centerline{\epsfxsize=17.0truecm \figinsert{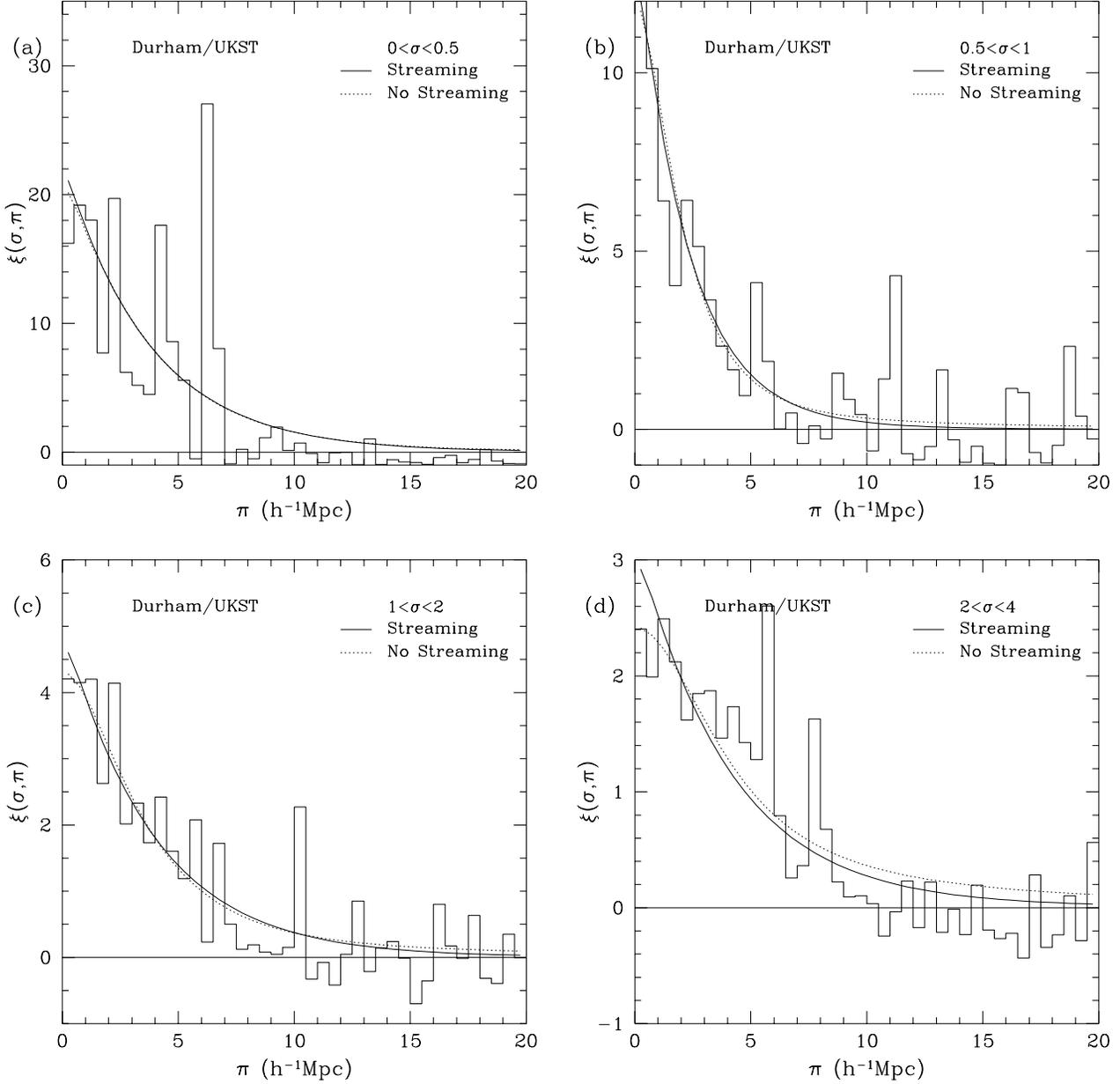}{0.0pt}}
\caption{Histograms of $\xi(\sigma,\pi)$ estimated from the Durham/UKST
survey as a function of $\pi$ for different (constant) values of
$\sigma$. The solid curve shows the minimum $\chi^{2}$ fit using the
modelling of Section~\ref{xispmodelsec} with a streaming model, while
the dotted curve shows the fit without the streaming model.}
\label{xipiukstfig}
\end{figure*}

We now apply the modelling of Section~\ref{xispmodelsec} to the
Durham/UKST survey. We use the optimally estimated $\xi(\sigma,\pi)$
data which was shown in Fig.~\ref{xispdurfig} and described in
Section~\ref{durxispsec}. Fig.~\ref{xipiukstfig} shows the histograms of
$\xi(\sigma,\pi)$ as a function of $\pi$ for different (constant)
values of $\sigma$. The solid curve shows the minimum $\chi^{2}$ fit of
the model from Section~\ref{xispmodelsec} with the streaming model. The
dotted curve shows the same fit without the streaming model. The
results of these fits are given in Table~\ref{xipiuksttab} where we
have assumed $\gamma = 1.7$. Small changes in this value of $\gamma$ do
not significantly affect our results. We use the standard deviations on
an individual LCDM mock catalogue in the $\chi^{2}$ fits. These fits
are more than likely biased low by the non-independent nature of the
points. The errors on the parameters quoted in Table~\ref{xipiuksttab}
come from the appropriate $\Delta\chi^{2}$ contour about the minimum.

We can make a few comments about these results. Firstly, the streaming
model is again required to produce consistent fits for $\veldisp$ for
$\sigma >$ 1-2$h^{-1}$Mpc (assuming that it does not change with
$\sigma$). Secondly, although not shown we also fitted for a Gaussian
distribution function (rather than the exponential one favoured by the
simulations) and found that the noise in the data implied a similar
quality of fit for both forms. Therefore, this data did not allow one
to distinguish between the two models. Finally, we combine the results
for the exponential velocity distribution function with the streaming
model at each $\sigma$ value to produce an overall estimate of
$\veldisp$ and $r_{0}$. Assuming that each quoted estimate is
independent gives $\veldisp = 416 \pm 36$ kms$^{-1}$, $r_{0} = 4.6 \pm
0.2h^{-1}$Mpc. We make the comment that this value of $r_{0}$ agrees
well with those estimated from the Durham/UKST survey using other
methods, see Section~\ref{realreddissec}. These results are discussed
in more detail in Section~\ref{discusssec} where comparisons with the
results from structure formation models and other redshift surveys are
made.

\begin{table}
\begin{center}
\caption{Minimum $\chi^{2}$ fits for $\veldisp$ and $r_{0}$ from the
Durham/UKST survey using the exponential distribution function with and
without the streaming model.} \label{xipiuksttab}
\begin{tabular}{cccc}
{\large $\sigma$} & {\large $\veldisp$} & {\large $r_{0}$} & {\large
$\chi^{2}$} \\
{\large ($h^{-1}$Mpc)} & {\large (kms$^{-1}$)} & {\large ($h^{-1}$Mpc)}
& {\large (N$_{bin} = 40$)} \\
& & & \\
\multicolumn{4}{c}{{\large Durham/UKST}} \\
\multicolumn{4}{c}{{\large Streaming Model}} \\
$[0,0.5]$ & $510 \pm 120$ & $5.1 \pm 0.6$ & 19.0 \\
$[0.5,1]$ & $300 \pm 50$ & $4.7 \pm 0.3$ & 23.5 \\
$[1,2]$ & $500 \pm 65$ & $4.5 \pm 0.2$ & 29.6 \\
$[2,4]$ & $500 \pm 65$ & $4.7 \pm 0.4$ & 48.0 \\
\multicolumn{4}{c}{{\large No Streaming Model}} \\
$[0,0.5]$ & $470 \pm 130$ & $5.2 \pm 0.6$ & 19.1 \\
$[0.5,1]$ & $180 \pm 70$ & $4.8 \pm 0.4$ & 24.0 \\
$[1,2]$ & $270 \pm 90$ & $4.8 \pm 0.3$ & 29.5 \\
$[2,4]$ & $180 \pm 80$ & $5.5 \pm 0.2$ & 56.1 \\
\end{tabular}
\end{center}
\end{table}

\section{Linear Scales: Infall and $\Omega^{0.6}/b$} \label{linsec}

We now use modelling techniques developed from the linear theory result
of Kaiser (1987) to estimate the quantity $\Omega^{0.6}/b$ from the
Durham/UKST survey. We saw in Section~\ref{xispsec} that on non-linear
scales the velocity dispersion was mainly responsible for the
anisotropies in $\xi(\sigma,\pi)$. However, in Section~\ref{nonlinsec}
we saw that to produce consistent results it was also necessary to
incorporate a model which imitated the streaming motions of galaxies.
It is this infall of galaxies into overdense regions that visibly
distorts the $\xi(\sigma,\pi)$ contours on larger scales and this
compression in the line of sight direction is now quantitatively
measured.

\subsection{Modelling $\xi(\sigma,\pi)$} \label{xisplinmodsec}

Kaiser (1987) used the distant observer approximation in the linear
regime of gravitational instability to show that the strength of an
individual plane wave as measured in redshift space is amplified over
that measured in real space by a factor
\begin{equation}
\delta_{\bf k}^{s} = \delta_{\bf k}^{r} \left( 1 + \mu_{\bf kl}^{2}
f(\Omega)/b \right) , \label{deltars}
\end{equation}
where $\delta_{\bf k}^{r}$ and $\delta_{\bf k}^{s}$ are the Fourier
amplitudes in real ($r$) and redshift ($s$) space, respectively,
$\mu_{\bf kl}$ is the cosine of the angle between the wavevector (${\bf
k}$) and the line of sight (${\bf l}$), $f(\Omega) \simeq \Omega^{0.6}$
is the logarithmic derivative of the fluctuation growth rate (e.g.
Peebles 1980) and $b$ is the linear bias factor relating the galaxy and
mass distributions, $(\Delta\rho/\rho)_{g} = b (\Delta\rho/\rho)_{m}$.
The distant observer approximation restricts use of
equation~\ref{deltars} to opening angles in a redshift survey of less
than $\sim$$50^{\circ}$ which can cause a systematic effect at the
$\sim$$5\%$ level in $f(\Omega)/b$ (Cole, Fisher \& Weinberg 1994).
This ($1 + \mu_{\bf kl}^{2} f(\Omega)/b$) factor propagates directly
through to the power spectrum, $P(k,\mu_{\bf kl}) \equiv \left\langle
\delta_{\bf k} \delta_{\bf k}^{*} \right\rangle$
\begin{equation}
P^{s}(k,\mu_{\bf kl}) = P^{r}(k) \left( 1 + \mu_{\bf kl}^{2}
f(\Omega)/b \right)^{2} , \label{pkbeta}
\end{equation}
where the real space $P^{r}(k)$ is assumed to be an isotropic function
of $k$ only. Thus the anisotropy is a strong function of the angle
between $\bf k$ and $\bf l$. Integrating over all $\mu_{\bf kl}$ gives
the angle-averaged redshift space $P^{s}(k)$
\begin{eqnarray}
P^{s}(k) & = & \frac{\int_{-1}^{1} d\mu_{\bf kl} P^{s}(k,\mu_{\bf
kl})}{\int_{-1}^{1} d\mu_{\bf kl}} , \\
& = & P^{r}(k) \left[ 1 + \frac{2}{3}\frac{f(\Omega)}{b} +
\frac{1}{5}\left( \frac{f(\Omega)}{b} \right)^{2} \right] .
\label{pkavgbeta}
\end{eqnarray}
Fourier transforming equation~\ref{pkavgbeta} (which has no explicit
$\mu_{\bf kl}$ dependence) gives the corresponding relation between the
angle-averaged $\xi(s)$ and $\xi(r)$
\begin{equation}
\xi(s) = \xi(r) \left[ 1 + \frac{2}{3}\frac{f(\Omega)}{b} +
\frac{1}{5}\left( \frac{f(\Omega)}{b} \right)^{2} \right] ,
\label{xiavgbeta}
\end{equation}
assuming that $\xi(r)$ is an isotropic function of $r$ only. If the
volume integral of $\xi$ is defined as
\begin{equation}
J_{3}(x) = \int_{0}^{x} \xi(y) y^{2} dy , \label{j3eqn}
\end{equation}
then
\begin{equation}
J_{3}(s) = J_{3}(r) \left[ 1 + \frac{2}{3}\frac{f(\Omega)}{b} +
\frac{1}{5}\left( \frac{f(\Omega)}{b} \right)^{2} \right] .
\label{j3avgbeta}
\end{equation}
We will use equations~\ref{xiavgbeta} and~\ref{j3avgbeta} as two of our
methods for estimating $\Omega^{0.6}/b$.

Hamilton (1992) has extended Kaiser's (1987) analysis by Fourier
transforming equation~\ref{pkbeta} and noting that the cosine factor in
Fourier space, $\mu_{\bf kl}^{2} \equiv k_{\bf l}^{2}/k^{2}$, becomes a
differential operator in real space, $(\partial/\partial {\bf
\left|l\right|})^{2} (\nabla^{2})^{-1}$. Therefore
equation~\ref{pkbeta} becomes
\begin{equation}
\xi^{s}(r,\mu_{\bf rl}) = \left( 1 + (f(\Omega)/b) (\partial/\partial
{\bf \left|l\right|})^{2} (\nabla^{2})^{-1} \right)^{2} \xi^{r}(r) ,
\end{equation}
where $\mu_{\bf rl}$ is the cosine of the angle between the pair
separation, $\bf r$, and the line of sight, $\bf l$. Hamilton (1992)
then shows that the solution of this equation can be written in terms
of the first three even spherical harmonics of $\xi^{s}(r,\mu_{\bf
rl})$, with the higher moments zero (and all odd moments zero by
definition)
\begin{equation}
\xi^{s}(r,\mu_{\bf rl}) = \xi_{0}(r) P_{0}(\mu_{\bf rl}) + \xi_{2}(r)
P_{2}(\mu_{\bf rl}) + \xi_{4}(r) P_{4}(\mu_{\bf rl}) ,
\end{equation}
where $P_{l}(\mu_{\bf rl})$ are the usual Legendre polynomials and
\begin{equation}
\xi_{l}(r) = \frac{2l + 1}{2} \int_{-1}^{1} \xi^{s}(r,\mu_{\bf rl})
P_{l}(\mu_{\bf rl}) d\mu_{\bf rl} . \label{sphhardefn}
\end{equation}
Arguably the most useful form of Hamilton's (1992) solution is that
expressed in terms of $\xi_{0}$ and $\xi_{2}$ only
\begin{eqnarray*}
& \left[ 1 + \frac{2}{3}\frac{f(\Omega)}{b} + \frac{1}{5} \left(
\frac{f(\Omega)}{b} \right)^{2} \right] \xi_{2}(r) = & \\
& \left[ \frac{4}{3}\frac{f(\Omega)}{b} + \frac{4}{7} \left(
\frac{f(\Omega)}{b} \right)^{2} \right] \left( \xi_{0}(r) -
\frac{3}{r^{3}} \int_{0}^{r} \xi_{0}(s) s^{2} ds \right) , &
\end{eqnarray*}
or by defining
\begin{eqnarray}
\tilde{\xi}_{0}(r) & = & - \xi_{0}(r) + \frac{3}{r^{3}} \int_{0}^{r}
\xi_{0}(s) s^{2} ds , \\
\tilde{\xi}_{2}(r) & = &  - \xi_{2}(r) ,
\end{eqnarray}
it can be written as
\begin{equation}
\frac{\tilde{\xi}_{2}}{\tilde{\xi}_{0}} = \frac{\left( \frac{4}{3}
\beta + \frac{4}{7} \beta^{2} \right)}{\left( 1 + \frac{2}{3} \beta +
\frac{1}{5} \beta^{2} \right)} . \label{xiharbeta}
\end{equation}
We will use the ratio in equation~\ref{xiharbeta} as our third method
for estimating $\Omega^{0.6}/b$.

\subsection{Testing the Modelling} \label{lincdmsec}

We have tested this modelling with the LCDM simulations and mock
catalogues. The SCDM model was not analysed because we felt that the
one-dimensional rms galaxy pairwise velocity dispersion too strongly
dominated the $\xi(\sigma,\pi)$ plots for a sensible answer to be
obtained. As will be shown below this is also the case for some aspects
of the LCDM model. Also, in Section~\ref{xisplinmodsec} we noted that
equation~\ref{deltars} and the subsequent analysis is only strictly
correct in the distant observer approximation and that data from
redshift survey opening angles of $\leq 50^{\circ}$ should only really
be used. For redshift surveys geometrically similar to the Durham/UKST
survey this restriction has a negligible impact on the analysis
techniques.

\begin{figure}
\centering
\centerline{\epsfxsize=8.5truecm \figinsert{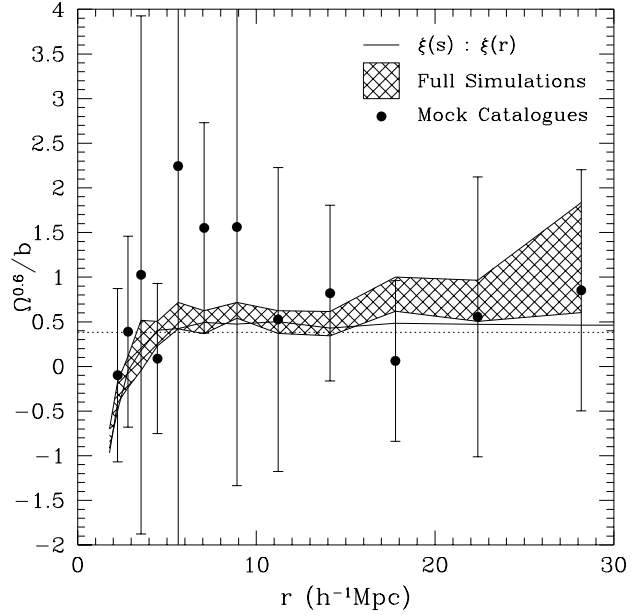}{0.0pt}}
\caption{Estimates of $\Omega^{0.6}/b$ obtained from the LCDM model
using the method involving the ratio of real to redshift space 2-point
correlation functions (equation~\ref{xiavgbeta}). The solid line
denotes the results obtained by averaging the $\xi$'s from each
$N$-body simulation and then manipulating to get an estimate of
$\Omega^{0.6}/b$. The shaded area denotes the $68\%$ confidence region
in $\Omega^{0.6}/b$ obtained from the scatter seen between the $N$-body
simulation $\xi$ results. The solid points denote the mean of the
results obtained from the mock catalogues. The error bars on these
points are the observed standard deviation on an individual mock
catalogue.} \label{betacdmxifig}
\end{figure}

\begin{figure}
\centering
\centerline{\epsfxsize=8.5truecm \figinsert{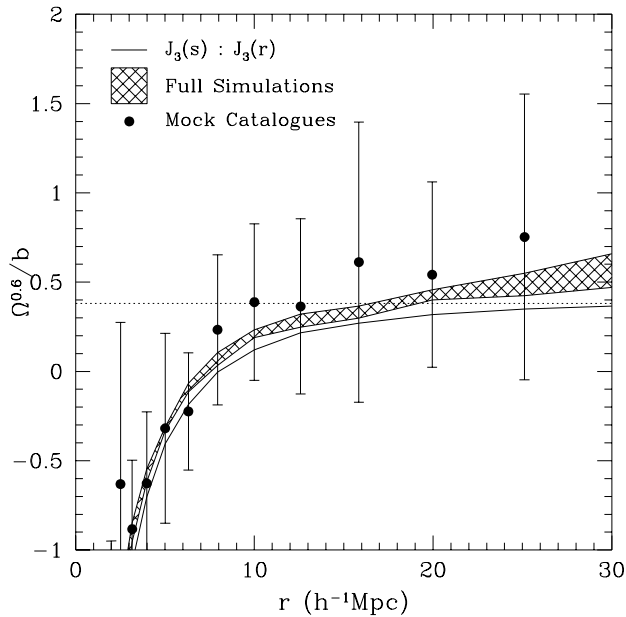}{0.0pt}}
\caption{The same as Fig.~\ref{betacdmxifig} but the $\Omega^{0.6}/b$
results use the method involving the ratio of real to redshift space
volume integrated 2-point correlation functions
(equation~\ref{j3avgbeta}).} \label{betacdmj3fig}
\end{figure}

\begin{figure}
\centering
\centerline{\epsfxsize=8.5truecm \figinsert{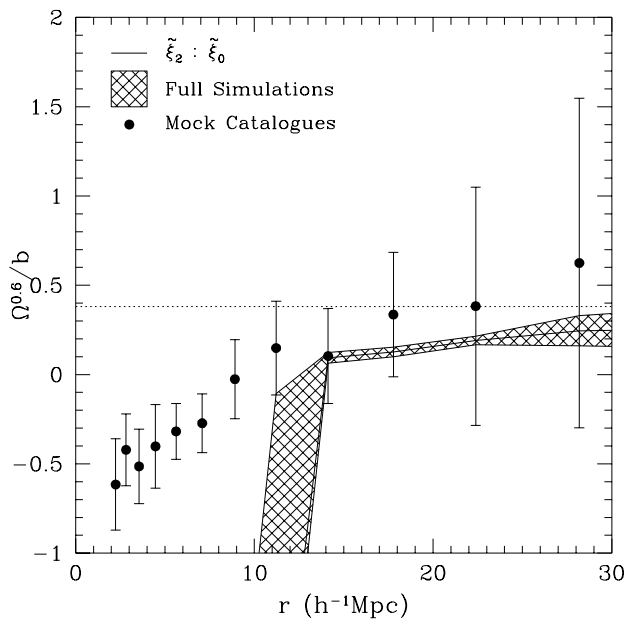}{0.0pt}}
\caption{The same as Fig.~\ref{betacdmxifig} but the $\Omega^{0.6}/b$
results use the method involving a ratio incorporating the second and
zeroth spherical harmonics of the redshift space 2-point correlation
function (equation~\ref{xiharbeta}).} \label{betacdmharfig}
\end{figure}

In Figs.~\ref{betacdmxifig},~\ref{betacdmj3fig} and~\ref{betacdmharfig}
the following conventions are adopted: the dotted line denotes the
expected value of $f(\Omega)/b \simeq \Omega^{0.6}/b = (0.2)^{0.6}/1
\simeq 0.38$; the solid line denotes the result for $\Omega^{0.6}/b$
obtained from the average of the LCDM simulations (i.e. estimate an
averaged $\xi(\sigma,\pi)$ from the 5 simulations and then manipulate
to get one value of $\Omega^{0.6}/b$); the shaded area denotes the
$68\%$ confidence region in $\Omega^{0.6}/b$ obtained from the scatter
seen between the LCDM simulations (i.e. take the $\xi(\sigma,\pi)$ from
each of the 5 simulations, manipulate to get 5 values of
$\Omega^{0.6}/b$ and then average at the end); and the points with
error bars are the mean and $1\sigma$ scatter seen in the LCDM mock
catalogues (i.e. take the $\xi(\sigma,\pi)$ from each of the 15 mock
catalogues, manipulate to get 15 values of $\Omega^{0.6}/b$ and then
average at the end). These errors are the observed standard deviation
between the mock catalogues and therefore reflect the error on an
individual mock catalogue.

The results from equation~\ref{xiavgbeta} are shown in
Fig.~\ref{betacdmxifig}. This method uses the ratio of the real to
redshift space $\xi$'s to estimate $\Omega^{0.6}/b$. We estimate
$\xi(s)$ using the methods described in Paper III. We estimate $\xi(r)$
using the Abel inversion of the projected correlation function,
$w_{v}(\sigma)$, with a $\pi_{cut}=30h^{-1}$Mpc, as described in
Section~\ref{xirsec}. Examining the results obtained from the averaged
$\xi$ of the simulations (solid line), it appears that the method
itself is not dominated by non-linear effects above $\sim$6$h^{-1}$Mpc,
although they could cause the $\sim$0.1 systematic offset in
$\Omega^{0.6}/b$ that is seen out to $> 30h^{-1}$Mpc.  Unfortunately,
the results obtained by averaging the 5 estimates of $\Omega^{0.6}/b$
from each (full) simulation show that noise begins to dominate the
inversion process between 15-20$h^{-1}$Mpc. For the mock catalogues
noise dominates at all scales and the results resemble a scatter plot
in places. Overall, this method appears relatively insensitive to
non-linear effects but the large scatter (for the mock catalogues)
renders the method almost useless in this case.

The results from equation~\ref{j3avgbeta} are shown in
Fig.~\ref{betacdmj3fig}. This method uses the ratio of the real to
redshift space $J_{3}$'s to estimate $\Omega^{0.6}/b$. We estimate
$J_{3}(s)$ by evaluating the integral in equation~\ref{j3eqn} using the
above $\xi(s)$. We estimate $J_{3}(r)$ from the same integral but using
the above $\xi(r)$. Examining the results obtained from the averaged
$\xi$ of the simulations, it appears that the method itself is not
dominated by non-linear effects above $\sim$15$h^{-1}$Mpc. The results
obtained by averaging the 5 estimates of $\Omega^{0.6}/b$ from each
simulation are similarly consistent with the expected value. The
results obtained from the mock catalogues also reproduce the correct
answer but with a larger scatter. Overall, this method is considerably
more successful (in terms of more accurate results) than the
$\xi(s)/\xi(r)$ one. However, one does pay a price by having to go to
larger scales before the non-linearities become negligible and the
integration procedure also makes the points highly correlated.

The results from equation~\ref{xiharbeta} are shown in
Fig.~\ref{betacdmharfig}. This method uses the ratio involving the
second and zeroth spherical harmonics of $\xi$ to estimate
$\Omega^{0.6}/b$. The $\xi_{l}$'s are estimated using
equation~\ref{sphhardefn}, which in practice becomes
\begin{equation}
\xi_{l}(r) = (2l + 1) \Delta\mu_{\bf rl} \sum_{i}^{(\mu^{i}_{\bf
rl}>0)} \xi^{s}(r,\mu^{i}_{\bf rl}) P_{l}(\mu^{i}_{\bf rl}) ,
\label{harmomcalc}
\end{equation}
and we use a binning of $\Delta\mu_{\bf rl} = 0.2$. Examining the
results obtained from the averaged $\xi$ of the simulations, it appears
that the method itself is severely affected by non-linearities which
cause a negative value of $\Omega^{0.6}/b$ to be measured until
$\sim$13$h^{-1}$Mpc. (Such a negative value is obviously unphysical and
simply due to the shape of the $\xi(\sigma,\pi)$ contours.) Similar
results are found by averaging the 5 estimates of $\Omega^{0.6}/b$ from
each simulation. The results obtained from the mock catalogues trace
the simulation results adequately, although not on scales $<
10h^{-1}$Mpc where they appear biased high. Overall, the impression is
that the non-linearities make this method redundant. However, when
considering the Durham/UKST survey one must remember that its value of
the velocity dispersion is almost half that of the LCDM model.
Therefore, the elongation in $\xi(\sigma,\pi)$ which causes the above
problems will be significantly lower.

\subsection{Results from the Durham/UKST Survey}

\begin{figure}
\centering
\centerline{\epsfxsize=8.5truecm \figinsert{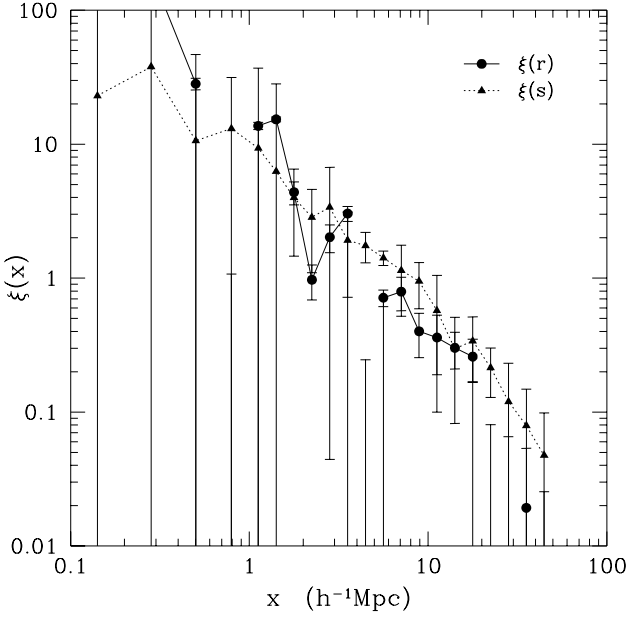}{0.0pt}}
\caption{Comparison of the real and redshift space 2-point correlation
functions optimally estimated from the Durham/UKST survey. $\xi(s)$ was
estimated directly from the redshift survey (solid triangles connected
by the dotted line) and $\xi(r)$ comes from the Abel inversion of
Section~\ref{abelsec} shown in Fig.~\ref{xirabelfig} (solid circles
connected by the solid line). The error bars denote the scatter seen
between in the 4 quadrants of the Durham/UKST survey.} \label{xidurfig}
\end{figure}

\begin{figure}
\centering
\centerline{\epsfxsize=8.5truecm \figinsert{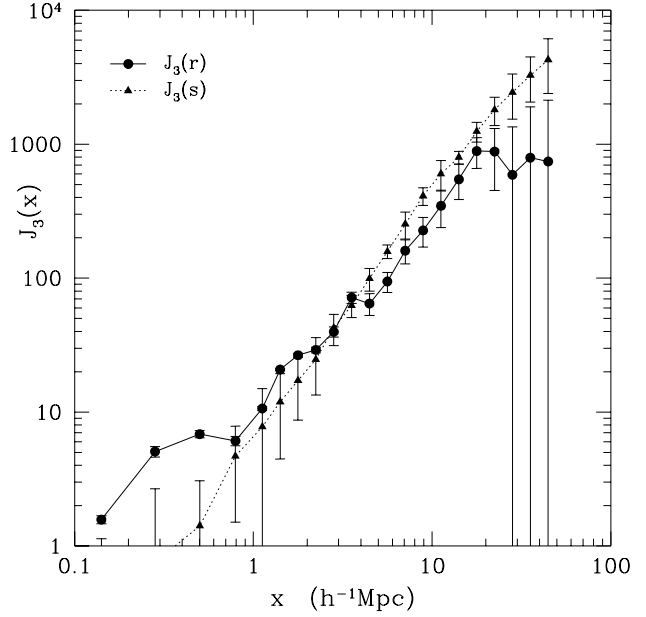}{0.0pt}}
\caption{Comparison of the real and redshift space $J_{3}$'s estimated
from the Durham/UKST survey. These were calculated using the integral
in equation~\ref{j3eqn} with the $\xi$'s of Fig.~\ref{xidurfig}. Again,
the error bars denote the scatter seen between in the 4 quadrants of
the Durham/UKST survey.} \label{j3durfig}
\end{figure}

\begin{figure}
\centering
\centerline{\epsfxsize=8.5truecm \figinsert{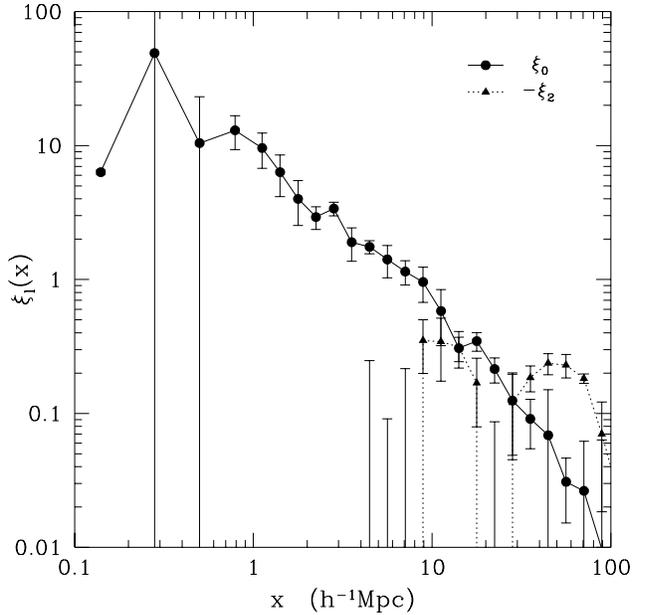}{0.0pt}}
\caption{The zeroth and second spherical harmonics from the Durham/UKST
survey. These were calculated using equation~\ref{harmomcalc} on the
optimally estimated $\xi^{s}(r,\mu_{\bf rl})$. The solid circles
connected by the solid line shows $\xi_{0}$, while the solid triangles
connected by the dotted line shows $-\xi_{2}$. Note that $-\xi_{2}$ is
plotted so that it is positive in the large scale region of interest.
Again, the error bars denote the scatter seen between in the 4
quadrants of the Durham/UKST survey.} \label{hardurfig}
\end{figure}

We now apply the modelling of Section~\ref{xisplinmodsec} to the
Durham/UKST survey. We calculate our results using the optimally
estimated 2-point correlation function described in
Section~\ref{durxispsec} and Paper III. Before presenting our results
of $\Omega^{0.6}/b$ we show the data which produced them.
Fig.~\ref{xidurfig} shows the real and redshift space $\xi$'s, where
$\xi(s)$ is estimated directly from the redshift survey and $\xi(r)$
comes from the Abel inversion of Section~\ref{abelsec} shown in
Fig.~\ref{xirabelfig}. Fig.~\ref{j3durfig} shows the real and redshift
space $J_{3}$'s, where the integral in equation~\ref{j3eqn} is applied
to the above $\xi$'s. Fig.~\ref{hardurfig} shows the zeroth and second
spherical harmonics estimated using equation~\ref{harmomcalc} on the
optimally estimated $\xi^{s}(r,\mu_{\bf rl})$. The error bars shown are
those estimated from the scatter between the 4 quadrants of the
Durham/UKST survey.  These errors were of a very similar size to those
estimated on an individual LCDM mock catalogue.

We comment briefly on the results in each of these figures. Firstly,
Fig.~\ref{xidurfig} shows that $\xi(r) > \xi(s)$ below
$\sim$1$h^{-1}$Mpc, while $\xi(s) > \xi(r)$ on larger scales. This was
described in Section~\ref{realreddissec}. Unfortunately, the noise in
these estimates is very high and it will more than likely make any
measurement of $\Omega^{0.6}/b$ redundant. Secondly,
Fig.~\ref{j3durfig} shows that, once again, at small separations
$J_{3}(r) > J_{3}(s)$, while at larger separations $J_{3}(s) >
J_{3}(r)$. There appears to be a near constant offset in the
real/redshift $\lg J_{3}$'s on scales 10-20$h^{-1}$Mpc (i.e. a constant
multiplicative factor in linear $J_{3}$). This should give a consistent
estimate of $\Omega^{0.6}/b$ on these scales. Finally,
Fig.~\ref{hardurfig} shows that $\xi_{2}$ is positive until
$\sim$8$h^{-1}$Mpc, which is caused by the elongation of the
$\xi(\sigma,\pi)$ contours parallel to the line of sight. On larger
separations $\xi_{2}$ is negative due to the compression of the
$\xi(\sigma,\pi)$ contours parallel to the line of sight. Therefore,
there appears to be a significant signal to measure in this method.

\begin{figure}
\centering
\centerline{\epsfxsize=8.5truecm \figinsert{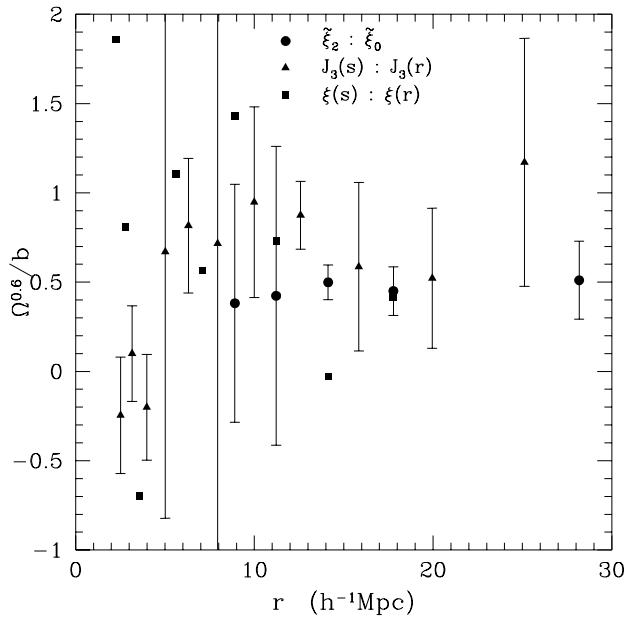}{0.0pt}}
\caption{Estimates of $\Omega^{0.6}/b$ obtained from the Durham/ UKST
survey. We use the three methods from equation~\ref{xiavgbeta} (solid
squares), equation~\ref{j3avgbeta} (solid triangles) and
equation~\ref{xiharbeta} (solid circles) and apply them to the data in
Figs.~\ref{xidurfig},~\ref{j3durfig} and~\ref{hardurfig},
respectively. The plotted errors denote the standard deviation seen
between the 4 quadrants of the Durham/UKST survey. For clarity, error
bars are not shown for the $\xi(s)/\xi(r)$ method because they are very
large and only cause confusion.} \label{betadurfig}
\end{figure}

We now present the results for $\Omega^{0.6}/b$ estimated from the
Durham/UKST survey. Fig.~\ref{betadurfig} shows the results of applying
equation~\ref{xiavgbeta} (solid squares), equation~\ref{j3avgbeta}
(solid triangles) and equation~\ref{xiharbeta} (solid circles) to the
data in Figs.~\ref{xidurfig},~\ref{j3durfig} and~\ref{hardurfig},
respectively. The plotted errors denote the standard deviation seen
between the 4 quadrants of the Durham/UKST survey. Similarly sized
errors were seen in the scatter between the individual LCDM mock
catalogues. For clarity, error bars are not shown for the
$\xi(s)/\xi(r)$ method because they are very large and only cause
confusion.

We can make a few comments about the results from these three methods.
Section~\ref{lincdmsec} showed that our region of interest is
$\sim$10-30$h^{-1}$Mpc because of non-linear effects on smaller scales
and noise on larger scales. For the $\xi(s)/\xi(r)$ method, the points
have no systematic trend (other than a large random scatter) and we
discount them from the further analysis. For the $J_{3}(s)/J_{3}(r)$
method, we obtain quite consistent results although one must remember
that these points are not independent because of the integration
prodedure. We choose to quote the value at $\sim$20$h^{-1}$Mpc as being
representative of our results, $\Omega^{0.6}/b = 0.52 \pm 0.39$. We
believe that we are being very conservative with this quoted error and,
for example, the point at $\sim$12$h^{-1}$Mpc only has an error of $\pm
0.19$. For the $\tilde{\xi}_{2}/\tilde{\xi}_{0}$ method, we again
obtain very consistent results and quote the value at
$\sim$18$h^{-1}$Mpc as being representative, $\Omega^{0.6}/b = 0.45 \pm
0.38$, where the quoted error is an average of the error bars in the
$\sim$10-30$h^{-1}$Mpc region and is therefore the typical error on any
{\it individual} point in this region. Naively combining the results in
a maximum likelihood manner in the $\sim$10-20$h^{-1}$Mpc region gives
an estimate of $\Omega^{0.6}/b = 0.48 \pm 0.11$. In this case the error
estimate is likely to be slightly underestimated given the generally
non-independent nature of $\xi$ points. These results are discussed in
more detail in Section~\ref{discusssec} where comparisons with the
results from structure formation models and other redshift surveys are
made.

\section{Discussion} \label{discusssec}

We now discuss the results obtained from the analysis presented in
Sections~\ref{nonlinsec} and~\ref{linsec}. Given the observed problems
with the unweighted estimate of $\xi$ (Paper III), we now favour the
weighted estimate of $\xi$ in all of our analysis. This explains any
slight numerical differences with respect to the quoted results of
Ratcliffe et al. (1996a). However, we state clearly that our overall
conclusions remain completely unchanged.

\subsection{The One-Dimensional RMS Pairwise Velocity Dispersion}

The minimum $\chi^{2}$ fit of an exponential distribution function
(with a streaming model) to the optimally estimated $\xi(\sigma,\pi)$
from the Durham/UKST survey gave a value for the one-dimensional rms
pairwise velocity dispersion of $416 \pm 36$ kms$^{-1}$. We note that
this quoted error is the formal error obtained from the $\chi^{2}$
statistic.  Therefore, is it likely to be slightly underestimated given
the non-independent nature of $\xi$. Our observed value is of
particular interest as recent estimates from new redshift surveys and
the re-analysis of old redshift surveys have been measuring larger
velocity dispersions than the canonical value of $340 \pm 40$
kms$^{-1}$ found by Davis \& Peebles (1983) from the CfA1 survey. For
example, using the CfA2/SSRS2 survey, Marzke et al. (1995) find $540
\pm 180$ kms$^{-1}$ and, using the Las Campanas survey, Lin et al.
(1996) find $452 \pm 60$ kms$^{-1}$. Also, Mo, Jing \& B\"{o}rner
(1993) measured large variations (200-1000 kms$^{-1}$) in the velocity
dispersion for a number of samples of similar size to CfA1 and they
show that the estimated velocity dispersion is sensitive to galaxy
sampling, especially dominant clusters the size of Coma. This new
estimate from the Durham/UKST survey is still on the slightly low side,
supporting the old Davis \& Peebles (1983) value, but is not
inconsistent ($>$3$\sigma$) with any of these other measured values.
When considering these values it is important to note that the
Durham/UKST survey covers a volume $\sim$$4 \times
10^{6}h^{-3}$Mpc$^{3}$, approximately twice that of the CfA2/SSRS2
survey and half that of the Las Campanas survey. Also, in an unbiased
(COBE-normalised) CDM model, Marzke et al. (1995) estimated that the
velocity dispersion would converge to 10\% within a volume $\sim$$5
\times 10^{6}h^{-3}$Mpc$^{3}$. Therefore, the measurement from the
Durham/UKST survey is hopefully both believable and representative of
the actual value in the Universe. Finally, we note that the Durham/UKST
survey does not contain any extremely dominant clusters (of Coma-like
size) and therefore will not be biased high by this.

The best estimate of the one-dimensional rms pairwise velocity
dispersion from the SCDM and LCDM simulations was 980 and 835
kms$^{-1}$, respectively, obtained using an exponential distribution
function. These estimates agree well with the actual value of the
velocity dispersion measured directly from the $N$-body simulations.
However, both of these values are inconsistent with the measured value
from the Durham/UKST survey at high levels of significance. In fact,
taking the most negative approach possible (i.e. using the individual
mock catalogue velocity dispersion error bar), one still finds a
significant rejection ($>$3$\sigma$) of both CDM models. However, it
should be noted that a significant velocity bias, $b_{v}$, between the
matter and galaxy velocity distributions ($b_{v}$$\sim$0.4), see
Couchman \& Carlberg (1992), would allow consistent results between the
models and the data. Also, this rejection of the CDM models assumes
that the simple models of linear biasing used here (Bardeen et al.
1986) are an adequate description of the galaxy formation process.

\subsection{Infall and $\Omega^{0.6}/b$}

The best estimates of $\Omega^{0.6}/b$ from the Durham/UKST survey are
$0.52 \pm 0.39$ for the $J_{3}(s)/J_{3}(r)$ method and $0.45 \pm 0.38$
for the $\tilde{\xi}_{2}/\tilde{\xi}_{0}$ method, where we have
(conservatively) quoted the estimate at one separation only. Given the
integration procedure involved in the $J_{3}$ method we do not attempt
to combine these points. However, naively combining the
$\tilde{\xi}_{2}/\tilde{\xi}_{0}$ estimates gives $\Omega^{0.6}/b =
0.48 \pm 0.11$. This error is likely to be slightly underestimated
given the non-independent nature of $\xi$. Our estimates can be compared
with other {\it optical} values of $\Omega^{0.6}/b$ estimated using
similar methods involving redshift space distortions. Peacock \& Dodds
(1994) use the real and redshift space power spectrum estimates of
various cluster, radio, optical and IRAS samples to measure
$\Omega^{0.6}/b = 0.77 \pm 0.16$. Loveday et al. (1996) use the $J_{3}$
method to measure $\Omega^{0.6}/b = 0.48 \pm 0.12$ for the APM-Stromlo
survey. Lin et al. (1996) use the $\tilde{\xi}_{2}/\tilde{\xi}_{0}$
method to measure $\Omega^{0.6}/b = 0.5 \pm 0.25$ for the Las Campanas
survey.

As one can see, all these observed values of $\Omega^{0.6}/b$ are
consistent with $\sim$0.5. Therefore, using two fiducial values
\mbox{for $b$}, $b=1$ implies and the universe is open,
$\Omega$$\sim$0.3, and $b$$\sim$2 implies the universe has the
critical-density, $\Omega=1$.

Finally, we state that our $\Omega^{0.6}/b$ estimates are consistent
with those from the two CDM models of structure formation considered
here (SCDM and LCDM). However, given that these models predict
$(\Omega^{0.6}/b)$$\sim$0.4-0.6 we are unable to distinguish between
them.

\section{Conclusions} \label{concsec}

We have investigated the redshift space distortions in the Durham/UKST
Galaxy Redshift Survey using the 2-point correlation function
perpendicular and parallel to the line of sight, $\xi(\sigma,\pi)$. On
small, non-linear scales we observe an elongation of the $\xi$ contours
in the line of sight direction, which is due to the velocity dispersion
of galaxies in virial regions. This is the common ``Finger of God''
effect seen in redshift surveys. On larger, linear scales we observe a
compression of the $\xi$ contours in the line of sight direction, which
is due to the infall of galaxies into overdense regions.

We attempt to estimate the real space 2-point correlation function by
direct inversion of the projected correlation function. We use two
different inversion methods, Abel inversion and a new application of
the Richardson-Lucy technique. We have tested these methods on mock
catalogues drawn from cold dark matter (CDM) $N$-body simulations and
find that they reproduce the correct answer. We apply the methods to
the Durham/UKST survey and estimate $\xi(r)$. We find that a simple
power law model gives best fit real space parameters of the correlation
length, $r_{0} = 4.8 \pm 0.5 h^{-1}$Mpc for the Abel method and $r_{0}
= 4.6 \pm 0.3 h^{-1}$Mpc for the Richardson-Lucy method, and slope,
$\gamma = 1.6 \pm 0.3$ for the Abel method and $\gamma = 1.6 \pm 0.1$
for the Richardson-Lucy method. Our estimate is consistent with those
from other redshift surveys and with the inversion of the APM
$w(\theta)$ by Baugh (1996). 

We use standard modelling techniques (e.g. Peebles 1980) to estimate
the one-dimensional rms pairwise velocity dispersion of galaxies in the
Durham/UKST survey. These methods (which were again tested on the mock
catalogues) give an estimate of $\veldisp = 416 \pm 36$ kms$^{-1}$ on
$\sim$1$h^{-1}$Mpc scales. This value agrees well with recent estimates
from new redshift surveys and re-analysis of old redshift surveys
(Markze et al. 1995; Lin et al. 1996; Mo et al. 1993), although our
value is still consistent with the canonical value of Davis \& Peebles
(1983). We compare with the predictions of the standard CDM model
(SCDM; $\Omega h = 0.5$ \& $b=1.6$) and the low density CDM model with
a non-zero cosmological constant to ensure spatial flatness (LCDM;
$\Omega h = 0.2$, $\Lambda = 0.8$ \& $b=1.0$). We find that our results
are significantly ($>$3$\sigma$) below the estimates from these models
(assuming that linear biasing applies) and therefore these models are
inconsistent with the data in this context. This modelling also
produces an estimate of the real space correlation length from the
Durham/UKST survey of $r_{0} = 4.6 \pm 0.2 h^{-1}$Mpc, which is very
consistent with the previously quoted values.

We use the modelling techniques of Kaiser (1987) and Hamilton (1992) to
estimate $\Omega^{0.6}/b$ from the Durham/UKST survey. We test the
methods with the mock catalogues and find that consistent results can
be obtained. The Durham/UKST survey results from the $\xi(s)/\xi(r)$
method are too noisy to obtain a useful answer. The $J_{3}(s)/J_{3}(r)$
method produces an estimate of $\Omega^{0.6}/b = 0.52 \pm 0.39$, where
we quote the estimate (and error) at one point only
($\sim$20$h^{-1}$Mpc) due to the non-independent nature of the
integration procedure. The $\tilde{\xi}_{2}/\tilde{\xi}_{0}$ method
gives $\Omega^{0.6}/b = 0.45 \pm 0.38$ at $\sim$18$h^{-1}$Mpc, which is
representative of the results on these scales. Naively combining the
points in the range $\sim$10-20$h^{-1}$Mpc gives a maximum likelihood
fit of $0.48 \pm 0.11$, where the error is likely to be slightly
underestimated due to the general correlated nature of $\xi$ points.
A comparison with other optical estimates of $\Omega^{0.6}/b$ from
redshift space distortion methods gives very consistent results with a
value of $(\Omega^{0.6}/b)$$\simeq$0.5. This argues against an unbiased
critical-density universe ($b=1$ \& $\Omega=1$), instead favouring
either an unbiased low density universe ($b=1$ \& $\Omega$$\sim$0.3) or
a biased critical-density universe ($b$$\sim$2 \& $\Omega=1$). Also,
given that both CDM models considered here predict
$(\Omega^{0.6}/b)$$\sim$0.4-0.6 we cannot use our $\Omega^{0.6}/b$
results to distinguish between them.

Overall, combining these results with those presented in Paper III, we
find that the standard CDM model underpredicts our 2-point correlation
function results at large scales and overpredicts the one-dimensional
pairwise velocity dispersion at small scales.  Therefore, our results
argue for a model with a density perturbation spectrum more skewed
towards large scales, such as a low $\Omega$ CDM model with a
cosmological constant.

\section*{acknowledgments}

We are grateful to the staff at the UKST and AAO for their assistance
in the gathering of the observations. S.M. Cole, C.M. Baugh and V.R.
Eke are thanked for useful discussions and supplying the CDM
simulations. AR acknowledges the receipt of a PPARC Research
Studentship and PPARC are also thanked for allocating the observing
time via PATT and for the use of the STARLINK computer facilities.

\end{document}